%% file: main.tex
\documentclass[10pt,journal,compsoc]{IEEEtran}
\ifCLASSOPTIONcompsoc
\ifCLASSINFOpdf
\usepackage[pdftex]{graphicx}
\graphicspath{{figures/}{pictures/}{images/}{./}} 
\DeclareGraphicsExtensions{.pdf,.png,.jpg,.jpeg}
\else
\fi

\input{setup.tex}

\title{\toole: Narration-Centric Design of Animated Data Videos}

\graphicspath{{figures/}{pictures/}{images/}{./}} 

\usepackage{caption}
\usepackage{amsmath}
\usepackage{algorithm}
\renewcommand{\figurename}{Figure}
\usepackage{color}
\usepackage{booktabs}
\usepackage{paralist}
\usepackage{enumitem}
\usepackage[hidelinks]{hyperref}

\usepackage{scalerel}
\usepackage{tikz}
\usetikzlibrary{svg.path}

\definecolor{orcidlogocol}{HTML}{A6CE39}
\tikzset{
  orcidlogo/.pic={
    \fill[orcidlogocol] svg{M256,128c0,70.7-57.3,128-128,128C57.3,256,0,198.7,0,128C0,57.3,57.3,0,128,0C198.7,0,256,57.3,256,128z};
    \fill[white] svg{M86.3,186.2H70.9V79.1h15.4v48.4V186.2z}
                 svg{M108.9,79.1h41.6c39.6,0,57,28.3,57,53.6c0,27.5-21.5,53.6-56.8,53.6h-41.8V79.1z M124.3,172.4h24.5c34.9,0,42.9-26.5,42.9-39.7c0-21.5-13.7-39.7-43.7-39.7h-23.7V172.4z}
                 svg{M88.7,56.8c0,5.5-4.5,10.1-10.1,10.1c-5.6,0-10.1-4.6-10.1-10.1c0-5.6,4.5-10.1,10.1-10.1C84.2,46.7,88.7,51.3,88.7,56.8z};
  }
}

\newcommand\orcidicon[1]{\href{https://orcid.org/#1}{\mbox{\scalerel*{
\begin{tikzpicture}[yscale=-1,transform shape]
\pic{orcidlogo};
\end{tikzpicture}
}{|}}}}

\newcommand{\bpstart}[1]{\vspace{1mm} \noindent{\textbf{#1.}}}

\begin{document}
\author{
Yun Wang*~\orcidicon{0000-0003-0468-4043}, Leixian Shen*~\orcidicon{0000-0003-1084-4912}, Zhengxin You~\orcidicon{0000-0002-6152-3513}, Xinhuan Shu~\orcidicon{0000-0002-9736-4454}, \\Bongshin Lee~\orcidicon{0000-0002-4217-627X}, John Thompson~\orcidicon{0000-0002-3102-4035}, Haidong Zhang~\orcidicon{0000-0001-7530-9553}, and Dongmei Zhang~\orcidicon{0000-0002-9230-2799}

	\IEEEcompsocitemizethanks{
		\IEEEcompsocthanksitem 
 Y. Wang, H. Zhang, D. Zhang are with Microsoft.
 E-mail: $\{wangyun, haizhang, dongmeiz\}$@microsoft.com. \\Y. Wang is the corresponding author.
\IEEEcompsocthanksitem
L. Shen is with The Hong Kong University of Science and Technology.
E-mail: lshenaj@connect.ust.hk.
\IEEEcompsocthanksitem
Z. You is with The Hong Kong Polytechnic University.
E-mail: zhengxin.you@connect.polyu.hk.
\IEEEcompsocthanksitem
X. Shu is with Newcastle University.
E-mail: xinhuan.shu@gmail.com.
\IEEEcompsocthanksitem
B. Lee is with Yonsei University. 
E-mail: b.lee@yonsei.ac.kr.
\IEEEcompsocthanksitem
J. Thompson is with Autodesk.
E-mail: john.roger.thompson@gmail.com.

\IEEEcompsocthanksitem
Work done during L. Shen and Z. You’s internship at Microsoft, and when B. Lee and J. Thompson were with Microsoft Research.
\IEEEcompsocthanksitem
* These authors contributed equally to this work.
	}
	\thanks{Manuscript received XX XX, 2024; revised XX XX, 2024.}}

\markboth{\MakeUppercase{IEEE TRANSACTIONS ON VISUALIZATION AND COMPUTER GRAPHICS},~Vol.~XX, No.~X, XX~2024}%
{Shell \MakeLowercase{\textit{\textit{et al.}}}: Bare Demo of IEEEtran.cls for Computer Society Journals}

\IEEEtitleabstractindextext{

	\begin{abstract}
		\raggedright
        \input{sections/00-abstract}

	\end{abstract}
	
}
\maketitle
\IEEEdisplaynontitleabstractindextext
\IEEEpeerreviewmaketitle

\input{sections/01-introduction}
\input{sections/02-related-work}

\input{sections/03-system-design}
\input{sections/04-system}
\input{sections/05-examples}

\input{sections/06-user-study}
\input{sections/07-discussions}

\input{sections/08-conclusion}

\bibliographystyle{abbrv-doi-hyperref}

\bibliography{main}

\newpage
\begin{IEEEbiography}[{\includegraphics[width=1in,height=1.2in,clip,keepaspectratio]{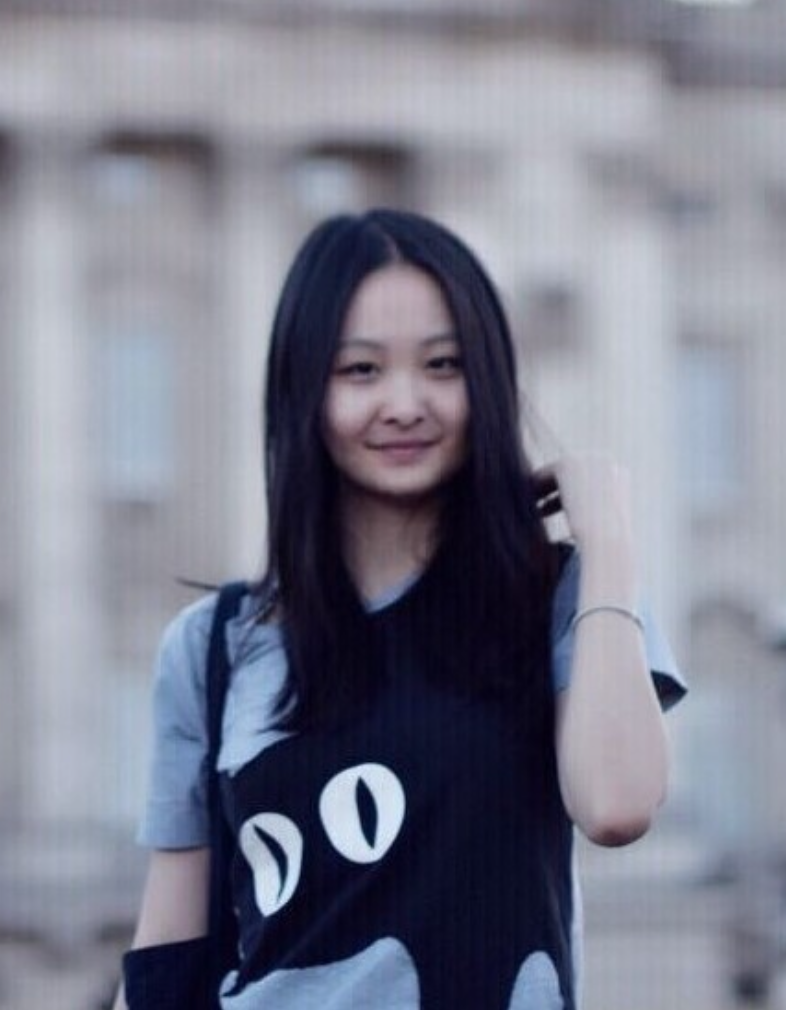}}]{Yun Wang}
is a senior researcher in the Data, Knowledge, Intelligence (DKI) Area at Microsoft. Her research lies in the intersection of Human-Computer Interaction, and Information Visualization. She develops innovative techniques and interactive systems to facilitate Human-AI Collaboration, Human-Data Interaction, Visual Communication, and Data Storytelling through an interdisciplinary approach. She received her Ph.D. from the Hong Kong University of Science and Technology. 
\end{IEEEbiography}
\vspace{-1.3cm}

\begin{IEEEbiography}[{\includegraphics[width=1in,clip]{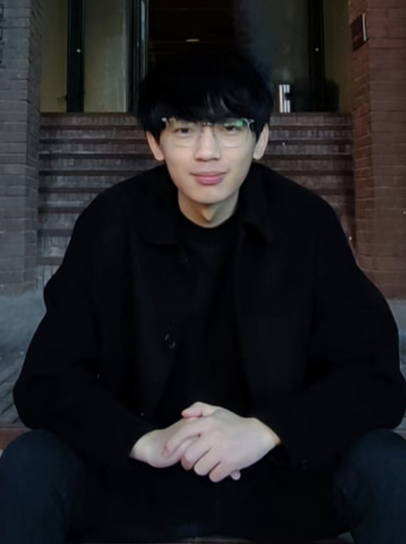}}]{Leixian Shen}
is a PhD student in the Department of Computer Science and Engineering at The Hong Kong University of Science and Technology. He received his master's degree in Software Engineering from Tsinghua University in 2023 and obtained his bachelor's degree in Software Engineering from Nanjing University of Posts and Telecommunications in 2020. His research interests include visual data analysis and storytelling.	
\end{IEEEbiography}
\vspace{-1.3cm}

\begin{IEEEbiography}[{\includegraphics[width=1in,height=1.2in,clip,keepaspectratio]{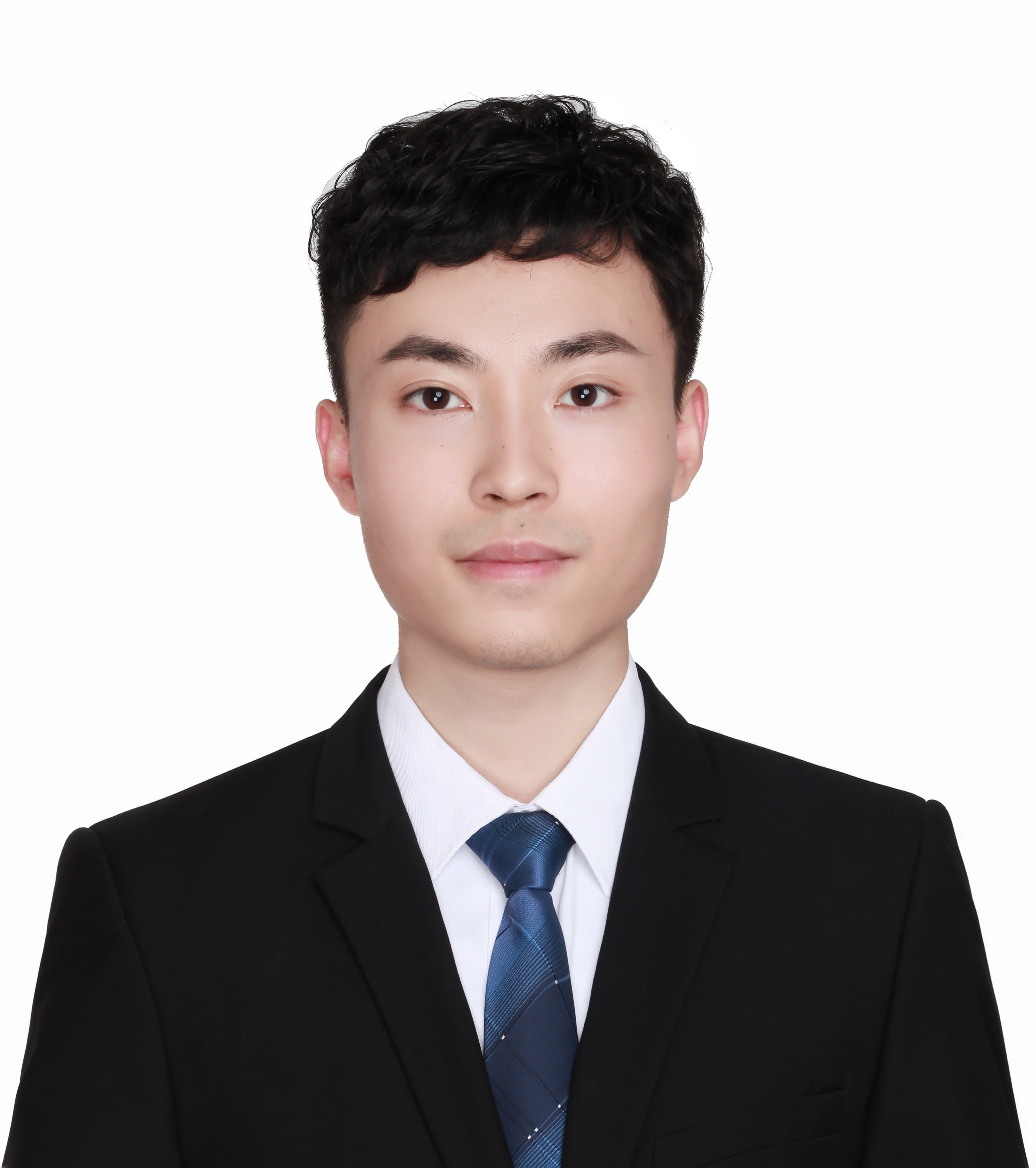}}]{Zhengxin You}
is currently a collaborative PhD student at the Hong Kong Polytechnic University and the Southern University of Science and Technology, China. His research interest is visualization for databases. Before this, he completed his undergraduate studies at the Southern University of Science and Technology. His work aims to enhance the understanding and usability of complex databases through innovative visualization techniques.
\end{IEEEbiography}
\vspace{-1.3cm}
\begin{IEEEbiography}[{\includegraphics[width=1in,height=1.2in,clip,keepaspectratio]{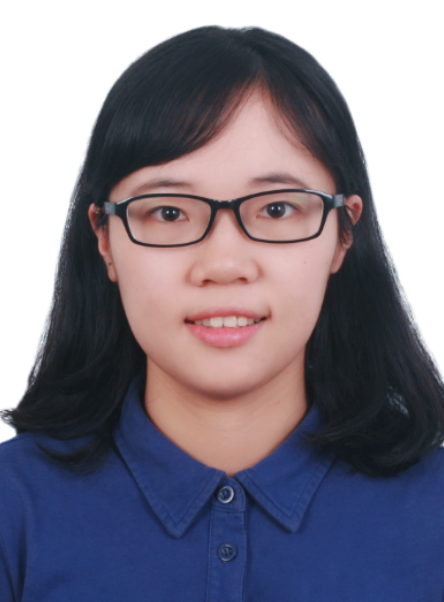}}]{Xinhuan Shu}
is currently a lecturer at Newcastle University. Her research focuses on designing expressive visualization techniques and human-AI interfaces that facilitate human-data interaction at various data activities, including data transformation, analysis, and storytelling. Prior that, she was a postdoctoral researcher at the University of Edinburgh and received her Ph.D. degree from HKUST. Her personal website is https://shuxinhuan.github.io/.
\end{IEEEbiography}
\vspace{-1.3cm}
\begin{IEEEbiography}[{\includegraphics[width=1in,height=1.2in,clip,keepaspectratio]{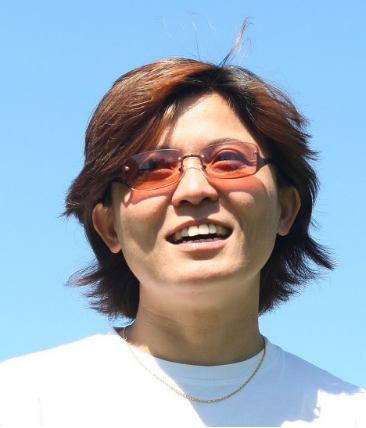}}]{Bongshin Lee} is a professor at Yonsei University. She conducts research on human-data interaction and human-computer interaction, with an overarching goal to empower everyone to achieve their goals by leveraging data, visualization, and technological advancements. Lee explores innovative ways to help people interact with data, by supporting data collection, reflection, and analysis, as well as data-driven communication. She received her PhD from the University of Maryland at College Park.
\end{IEEEbiography}
\vspace{-1.3cm}
\begin{IEEEbiography}[{\includegraphics[width=1in,height=1.2in,clip,keepaspectratio]{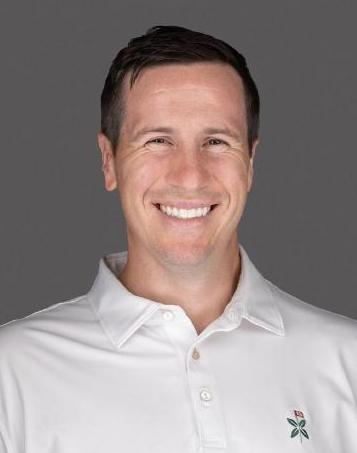}}]{John Thompson}
is a postdoctoral researcher at AutoDesk. He holds a Ph.D. degree and a master's degree from the Georgia Institute of Technology, as well as a B.E. degree from the University of Virginia. He was a postdoctoral researcher at Microsoft Research. His research interests include information visualization and data storytelling.
\end{IEEEbiography}
\vspace{-1.3cm}
\begin{IEEEbiography}[{\includegraphics[width=1.1in,height=1.2in,clip,keepaspectratio]{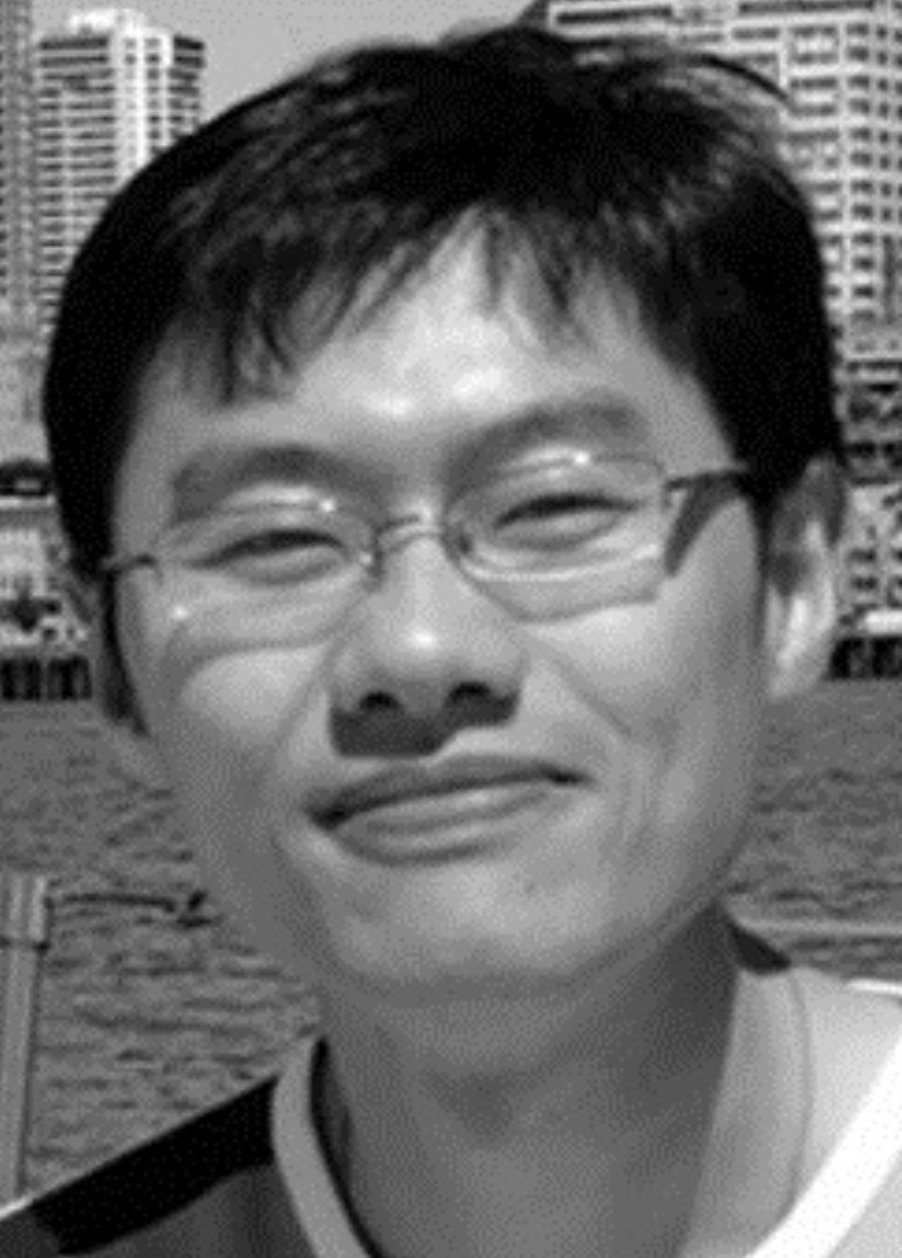}}]{Haidong Zhang}
is currently a principal architect with Microsoft. He received a Ph.D. degree in computer science from Peking University, China. His research interests include visualization and human-computer interaction.
\end{IEEEbiography}
\vspace{-1.3cm}
\begin{IEEEbiography}[{\includegraphics[width=1.1in,height=1.2in,clip,keepaspectratio]{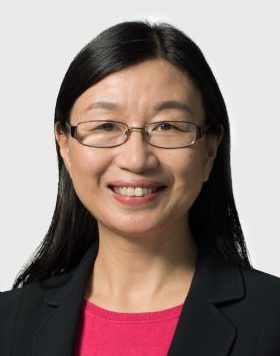}}]{Dongmei Zhang}
received the BE degree and ME degree from Tsinghua University, and the PhD degree in Robotics from the School of Computer Science at Carnegie Mellon University. She is a chief scientist at Microsoft, leading research in the area of Data, Knowledge, and Intelligence with research directions in data intelligence, knowledge computing, information visualization, and software engineering.
\end{IEEEbiography}

\clearpage

\end{document}

%% file: setup.tex
\usepackage{multicol}
\usepackage{multirow} 
\newcommand{\eg}{{\it e.g.,\ }}
\newcommand{\etal}{{\it et al.\ }}

\newcommand{\ie}{{\it i.e.,\ }}

\newcommand{\sutt}[1]{\textcolor{black}{ \textquotedblleft #1\textquotedblright}}
\newcommand{\tool}{WonderFlow~}
\newcommand{\toole}{WonderFlow}

\newcommand{\reviseminor}[1]{{\color{black} #1}}
\newcommand{\revise}[1]{{\color{black} #1}}



%% file: sections/00-abstract.tex
\revise{
Creating an animated data video with audio narration is a time-consuming and complex task that requires expertise. It involves designing complex animations, turning written scripts into audio narrations, and synchronizing visual changes with the narrations. This paper presents WonderFlow, an interactive authoring tool, that facilitates narration-centric design of animated data videos. WonderFlow allows authors to easily specify semantic links between text and the corresponding chart elements. Then it automatically generates audio narration by leveraging text-to-speech techniques and aligns the narration with an animation. WonderFlow provides a structure-aware animation library designed to ease chart animation creation, enabling authors to apply pre-designed animation effects to common visualization components. Additionally, authors can preview and refine their data videos within the same system, without having to switch between different creation tools. A series of evaluation results confirmed that WonderFlow is easy to use and simplifies the creation of data videos with narration-animation interplay.
}

%% file: sections/01-introduction.tex
\section{Introduction}


\IEEEPARstart{D}{ata} videos have gained increasing popularity in recent years, being widely used in digital journalism, knowledge sharing, business, etc. With data videos, 
data and numbers can be accurately described with charts and annotations \cite{amini2015understanding}, while insights and contexts behind the data are usually conveyed through voice narration \cite{wang2022investigating}. Data video creators leverage animations to demonstrate the data characteristics and echo them with voice narrations to attract viewers' attention and enhance cognition and memorability \cite{amini2018hooked}.

\reviseminor{Creating a data video involves various tasks, such as organizing data insights, recording narration audio, designing animation effects, aligning visual effects with the audio narration, and etc. These tasks require not only significant amounts of time and effort, but also expertise in multiple fields---data analysis, animation design, and audio \& video production---and professional tools \cite{xia2022millions}. 
Therefore, creating animated data videos enriched with audio narrations involves a complex workflow and is not accessible to people without the required expertise. 

Recently, several research efforts have been devoted to easing the workflow of creating data-driven animations. Researchers have adopted animation templates \cite{amini2017authoring, shi2021autoclips}, developed declarative grammar \cite{ge2020canis} and visual specification \cite{ge2021cast}, leveraged keyframes \cite{thompson2021data,ge2021cast}, and designed automated algorithms \cite{wang2021animated, kim2021gemini2} to facilitate animation authoring. 
Also, in addition to animation design, audio narration has been found to be an important aspect of visual storytelling that enhances cognition \cite{clark_dual_1991}. Cheng \etal \cite{wang2022investigating} recently investigated the roles of narrations and animations by analyzing the semantics expressed in data videos and found that animations and narrations are linked in most cases.
However, existing chart animation authoring tools lack support for coordinating audio narration and visual animations, requiring authors to use two separate tools.}

\revise{To address the gap, we first initiated a formative study to comprehend the common workflows involved in producing data videos and figure out the challenges encountered by designers while creating data videos. 
During the study, the participants frequently mentioned that they needed support for animation design for visualizations, temporal alignment of narration and animation, and real-time preview of the data video. Since a visualization usually involves a large number of visual elements, the creation process of animations for visualization is generally very tedious and cumbersome. The authors need to consider the structures and semantics of the visualizations. 
To create data videos with audio narration, designers need to design and synthesize animations first, export video clips and import the video clips into video production tools (\eg Adobe Premiere), and coordinate narration and animation through video and audio editing operations. While designing animations, designers cannot preview the final videos and adjust the narration. The separation of the tools hinders an iterative and exploratory design process. As a result, despite the advances in visualization animation authoring research, designers of data videos may still resort to low-level animation design systems (\eg Adobe After Effect) to achieve fine-tuned adjustment of narration-animation interplay for data videos.

To address these challenges, we present a novel pipeline (Figure~\ref{fig:pipeline}) with a text-visual linking paradigm and visualization structure-aware animation library for easy creation of data videos. It simplifies the coordination of narration and animation, especially for novice users. In detail, the proposed pipeline helps to establish semantic links between static text segments and visual elements that facilitate the temporal alignment of narration and animation. With these links in place, narrations and animations can be automatically synchronized, reducing the coordination effort. In addition, to facilitate the animation design process, we propose visualization structure-aware animation presets that enable flexible re-usage of component-level animation design patterns of data visualizations. 
We introduce \toole, an interactive authoring tool that implements the text-visual linking paradigm for easy creation of data videos with narration-animation interplay. \tool utilizes an animation synthesis module that applies structure-aware animation presets to chart components based on author preferences. Furthermore, it employs text-to-speech techniques to automatically generate audio narration from the text. With \toole, authors can preview animation designs through the design panel and refine their videos.


We demonstrate the expressiveness of our approach through a gallery of diverse examples, and evaluate \toole's usability and capability through a user study with 10 novices (\ie participants with little animation design background) and interviews with four experts. 
All the novice participants could complete the data video design tasks and enjoyed exploring animation designs with \toole. The experts also appreciated the benefits of the text-visual linking paradigm and visualization structure-aware animations. We also compare \tool against PowerPoint by counting the minimal set of interactions.
Finally, we discuss insights learned from our study and potential future work for creating data videos. 

The main contributions of this work are:
\begin{compactitem}
    \item 
    A formative study that gains an understanding of typical workflows involved in data video creation, as well as identifying the challenges faced by practitioners throughout the process of crafting such videos.
    \item 
    An interactive authoring system, \toole, that facilitates the easy creation of data videos with narration-animation interplay, which is equipped with a new text-visual linking interaction paradigm and a visualization structure-aware animation library.
    \item 
    Four forms of evaluation: an example gallery, a user study with novices, interviews with experts, and an interaction-count comparison against existing tools. The results demonstrated \toole's expressiveness and usability.
\end{compactitem} 
}

%% file: sections/02-related-work.tex
\begin{figure*}[t]
\centering
\includegraphics[width=0.75\linewidth]{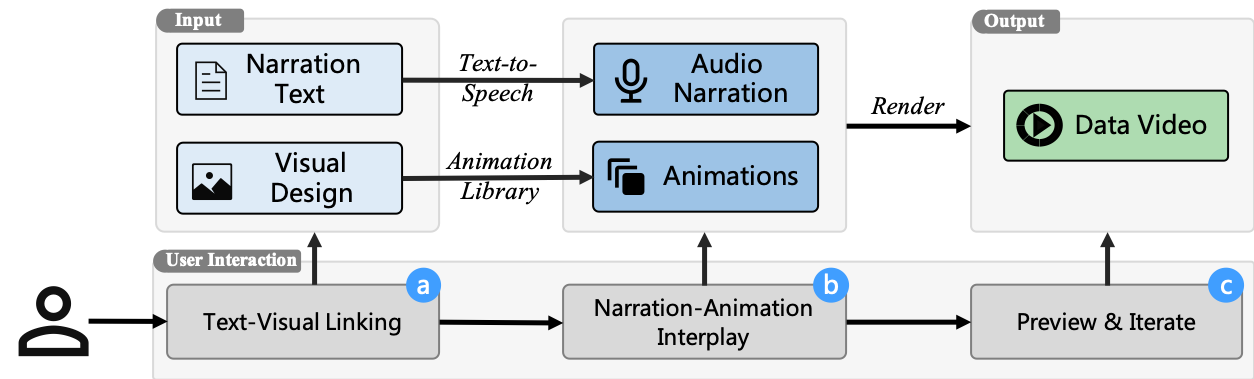}
\caption{
The pipeline of narration-centric design of animated data videos.
}
\label{fig:pipeline}
 \vspace{-10px}
\end{figure*}

\section{Related Work}
In this section, we discuss prior works from three perspectives---data videos, authoring tools and toolkits for data videos, and visualization-text interplay.

\subsection{Data Videos}
Data videos, which effectively combine both motion graphics and auditory stimuli to illustrate data~\cite{amini2015understanding}, are becoming an important medium in data-driven storytelling~\cite{segel2010narrative}. 
Viewers can easily follow the narratives while being engaged in attractive visuals and animations.
In recent years, a plethora of research has studied this genre in depth. 
For example, Amini \etal \cite{amini2015understanding} first systematically analyzed 50 data videos, from which they summarized the most commonly used visualization types and attention cues, and identified high-level narrative structures, \ie establisher, initial, peak, and release. 
They further validated that animations in data videos have a positive effect on viewer engagement~\cite{amini2018hooked}.
Shi \etal \cite{shi2021communicating} further examined the animated visual narratives in data videos and summarized a design space for motion design. 
Tang \etal \cite{tang2020narrative} proposed a taxonomy of narrative transitions in data videos, where they investigated how visual variables are used in transitions to build fluid narratives. 
Thompson \etal \cite{thompson2020understanding} identified design primitives of animated data graphics.
Besides, researchers have also worked on designing expressive animations to elevate data videos in terms of enhancing users' comprehension~\cite{waldner2014attractive, shu2020makes}, improving user engagement~\cite{li2020improving, wang2016animated, shu2021dancingwords}, and augmenting affective expressiveness~\cite{lan2021kineticharts}.


However, existing work mainly focused on visual and animation designs even though data videos often involve a close interplay of both audio narrations and visual animations to promote data story-telling. 
Very recently, Cheng \etal \cite{wang2022investigating} explored the roles of animations and narrations by analyzing 426 clips from 60 data videos and found a close link between animations and narrations. 
Furthermore, they identified four types of narration-animation relationships, \ie narrations and animations can echo each other by repeating, complementing additional information in animations, complementing additional information in narrations, and indirectly linking to each other. 
Despite the close link between narrations and animations, little work has explored the fine-grained integration of the two perspectives for bespoke data videos. Our research opens up a new research avenue towards flexible narration-centric design of animated data videos.


\subsection{Authoring Tools and Toolkits for Data Videos}
\revise{
The growing popularity of data videos and animated visualizations has in turn spurred the need for authoring tools and toolkits to facilitate the creation process.
Visualization toolkits (\eg $D3$~\cite{bostock2011d3}, gganimate~\cite{gganimate}, and Plotly~\cite{plotly}) allow programmers to create sophisticated animations. 
Animation-specific grammars (\eg Canis~\cite{ge2020canis}, Gemini~\cite{kim2021gemini,kim2021gemini2}, and Animated Vega-Lite~\cite{Zong2022}) propose high-level specifications to implement animated transitions of data graphics via a keyframe-based approach. 

Following the keyframe approach, interactive visualization authoring tools are developed to simplify the animation creation process by adding automation algorithms. For example, InfoMotion~\cite{wang2021animated} requires users to upload the final infographic figure and generate animation effect sequences automatically. 
Data Animator~\cite{thompson2021data} use keyframes of charts to facilitate the creation of animation by generating animated transitions between them leveraging the data attributes.
CAST~\cite{ge2021cast} is built on Canis~\cite{ge2020canis}, a declarative chart animation grammar that leverages data-enriched SVG charts to support auto-completion for constructing both keyframes and keyframe sequences of charts.
While these tools allow users to generate expressive animations for infographics and data graphics, users need to pre-define and manage keyframe sequences when creating data videos. 

Another group of interactive data video tools are developed to enable novices to create data animations. For example, Amini \etal \cite{amini2017authoring} proposed DataClips with a library of clips that convert data into pre-defined animations. 
DataClips enables analysts without knowledge of video design to choose clips from the library and concatenate them to produce data videos.
Shi \etal \cite{shi2021autoclips} then designed AutoClips, which improves the library with a set of fact-driven clips. 
Other recent tools focus on time-series data. Data2video~\cite{lu2020illustrating} and Roslingifier~\cite{Shin2022} study specific visual effects of animated scatterplots for temporal events, enabling the generation of scatterplot video for time-series data. 
These tools encapsulate animations with specific charts and develop a library of clips with fixed and finite combinations of animations and visualizations. 
However, clips of data videos in practice can be more flexible and complicated. Users are not able to control visualization components, such as axes and annotations with clip-level templates. 
In addition, users may adopt bespoke charts and annotation designs to describe the data insights. The chart elements in a clip can be animated diversely in response to different narrative content.
So clip-level presets are not able to support flexible integration of audio narrations and animations in creating data videos~\cite{wang2022investigating}. 

In this work, \tool supports a set of structure-aware animation presets, ranging from chart level, component level, to element level. It integrates narration and animation creation into one system from the visualization structure perspective. 
With \toole, users can directly construct text-visual links, and the tool automatically aligns the auto-generated audio narration and the user-selected animation effects, without burdening users with manually refining the narration-animation interplay on a timeline.}

\subsection{Visualization-Text Interplay}

There has been extensive research that explores the interplay of visualization and text~\cite{shen2022towards}. 
Empirical studies have shown the effectiveness of integrating text with visualizations in multiple aspects, including information recall improvement~\cite{zhi2019linking} and visualization comprehension~\cite{barral2020understanding,lalle2019gaze}. 
As such, many tools (\eg \cite{wang2019datashot,shi2021calliope, Sun2022}) generate templated textual explanations along with visualizations in data stories to facilitate communication.
Further, researchers have explored constructing text-chart links to achieve a more flexible interplay through direct manipulation, mixed-initiative, and automatic methods. 
For example, Sulatanum \etal \cite{sultanum2021leveraging} developed VizFlow, allowing users to specify text-chart links in different scrolly-telling layouts. 
To enhance the reading experience of data-driven documents, prior work has also explored building the reference between text with tables~\cite{badam2018elastic, kim2018facilitating} and charts~\cite{kong2014extracting} across the document. 
Kori~\cite{latif2021kori} provides a mixed-initiative method that uses NLP tools to construct interactive references between chart components and descriptive text. 
Lai \etal \cite{lai2020automatic} automatically generates an annotated visualization by synchronizing with textual description. 
Charagraph~\cite{Charagraph} enables users to interactively generate in-situ visualizations for real-time annotation of data-rich text.
DataParticles~\cite{DataParticles2023} enables users to interactively author a story narrative and its corresponding unit visualizations.
Our work is in line with the research to construct text-chart co-reference.
However, no existing research leverages text-visual linking to create data videos. Our work goes beyond text-visual linking to generate data videos that embody the text-visual links as the cohesive interplay between audio narration and visualization animations. 
In addition, we take charts in the SVG format to cater to the demand for expressive chart designs in data videos. 

%% file: sections/03-system-design.tex
\revise{
\section{Formative Study}
A formative study was undertaken to gather insights from expert data video designers and researchers. The primary objectives of this study were twofold: (1) to gain an understanding of the typical workflows involved in data video creation, and (2) to identify the challenges encountered by practitioners during the process of creating data videos.

\subsection{Participants}
To accomplish the aforementioned objectives, a total of 8 participants (3 males and 5 females) were recruited for the study, representing a diverse range of expertise from both academia and industry. The participants included professionals specializing in video design and visualization research. More demographic details are shown in Table \ref{tab: info}.
All of the designers possessed practical experience in crafting data videos using professional tools such as Adobe Premiere, and all visualization researchers have experience using simplified tools like Microsoft PowerPoint to create presentation videos. 
The recruitment involved a combination of word-of-mouth referrals and online advertising to ensure a diverse and representative participant pool.

\subsection{Procedure}
To begin, a retrospective analysis was conducted, where the participants were requested to share and showcase examples from their previous data storytelling projects. They were asked to walk through the creation process step by step and provide explanations for their design choices. 
Following the retrospective analysis, one-on-one semi-structured interviews were conducted with each participant. 
The participants were encouraged to recall additional details about their designs and discuss the challenges they encountered while crafting data videos, particularly in narration-animation interplay. Finally, the participants were prompted to share their expectations regarding future data video creation tools.
The entire process with each participant lasted about 60 minutes, and all sessions were recorded for subsequent analysis and reference purposes. 

\begin{table}
\centering
  \caption{Demographic information of our interviewees.}
  \setlength{\tabcolsep}{0.5mm}{
  \begin{tabular}{cccclc}
    \toprule
    \textbf{ID} & \textbf{Gender} & \textbf{Age} & \textbf{Education} & \textbf{Domain} & \textbf{Experience}\\
    \midrule
    E1 & Female & 24 & Master & Motion Graphics Designer & 4 years\\
    E2 & Female & 25 & Master & Animation Designer & 5 years\\
    E3 & Male & 31 & PhD & Visualization Researcher & 8 years\\
    E4 & Female & 26 & PhD & Media Arts Designer  & 5 years\\
    E5 & Female & 28 & PhD & Motion Graphics Designer & 7 years\\
    E6 & Male & 24 & Master & Visual Designer & 4 years\\
    E7 & Female & 25 & Master & Film Editor & 5 years\\
    E8 & Male & 32 & PhD & Visualization Researcher & 8 years\\
    \bottomrule
  \end{tabular}}
  \label{tab: info}
   \vspace{-10px}
\end{table}

\begin{table*}[htb]
\centering
  \caption{Tasks for creating data videos with narration-animation interplay.}
  \begin{tabular}{cll}
    \toprule
    \textbf{ID} & \textbf{Task} & \textbf{Task Description}\\
    \midrule
    T1 & Understand Narratives & 
    Identify key storylines and messages and analyze the script. \\
    T2 & Organize Visual Elements & 
    Group visual elements with semantic relationships to align with the narrative. \\
    T3 & Create Semantic References & 
    Build semantic links between the formulated narration segments and visual elements.\\
    T4 & Record Narration Audio & 
    Record clear and synchronized audio narration for the video.\\
    T5 & Design Visual Animations & 
    Design appropriate animations with semantics for the visual elements.\\
    T6 & Coordinate Audio and Animations & 
    Review audio and animation for synchronization using professional video editing tools.\\
    \bottomrule
  \end{tabular}
  \label{tab: step}
   \vspace{-10px}
\end{table*}

\subsection{Findings}
We summarized three key findings from the participants' feedback, \ie the common data video creation workflow, challenges in authoring data videos with narration-animation interplay, and design insights of the tool.

\subsubsection{Workflow for Data Video Creation}
All participants agreed that creating data videos is a complex, time-consuming, and experience-intensive process. This process usually involves six common tasks as shown in Table~\ref{tab: step}: understanding narratives, organizing visual elements, creating visual references, recording narration audio, designing visual animations, and coordinating audio and animations. These tasks are interconnected, and changes in one task (\eg refining the narrative text) can impact multiple other tasks (\eg audio recording and creating visual references). Therefore, the participants emphasized that the entire creation process is a trial-and-error and iterative refinement process. 
\reviseminor{Furthermore, among these tasks, T4 and T6 are specific to creating data videos with narration-animation interplay, while the other tasks are equally applicable to other forms of data storytelling such as scrollytelling and interactive data articles.}

\subsubsection{Challenges in Data Video Authoring}
Through the examination of cases, we engaged in discussions with the participants regarding the challenges they faced during the creation of narration-enriched data videos.

\textit{Inability to preview in real time.}
Creating data videos involves dynamic effects and the presentation of information, unlike static visual designs. Therefore, it is essential to enable a real-time preview of each segment to ensure alignment with the creator's intent.
However, existing tools developed for chart animation authoring lack support for coordinating audio narration and visual animation effects, requiring authors to use two separate tools. 
To create data videos with audio narration, authors need to synthesize and export video clips with animations first, import the video clips into video production tools (\eg Adobe Premiere), and coordinate narration and animation through video and audio editing operations.
Designers (E1, E4, and E7) emphasized that ``\textit{I cannot preview the final videos and adjust the narration while designing animations.}"
So the separation of the tools hinders an iterative and exploratory design process.

\textit{Inconvenience to align narration and animation.}
The transformation of static materials into dynamic presentations necessitates careful handling of temporal relationships. Once designers have prepared animations and narration audio and imported them into professional editing software, they often resort to manual time adjustment of keyframes. They had to repeatedly review the audio and animation to identify any synchronization discrepancies. Moreover, when encountering modifications in the narration, they had to update all synchronization information. 
A majority of the interviewees (6/8) expressed that this process was highly time-consuming and demanding.

\textit{Complexity of animation design for the visual components.}
Creating animations for visual components is a tedious process. It also requires expertise and experience to convey messages and express semantics vividly.
The keyframe-based approaches~\cite{thompson2021data,ge2021cast, wang2021animated} and declarative grammars~\cite{ge2020canis,kim2021gemini} give designers and researchers fine control over the properties of animations, but it requires users to have animation design or programming expertise and forces them to specify animation behaviors starting from scratch for every new visualization design. For novice and casual users, reusable animation templates can largely lower the learning curve~\cite{thompson2020understanding} and the required efforts. However, existing tools on animated visualization usually encapsulate animations into predefined video clips \cite{amini2017authoring, shi2021autoclips} and are not flexible for customized visualizations.

\subsubsection{Design of Authoring Tools}
Based on discussions with the participants regarding the limitations of the current data video workflow and their expectations for future tools, we have identified two common insights. The first insight involves integrating the entire workflow into a single tool that allows for convenient alignment of audio and animations, real-time video preview, and iterative adjustments \reviseminor{(DC1, DC2, DC4)}. To achieve this goal, most participants suggested that automatic Text-to-Speech technologies can be adopted for real-time narration updating instead of repeated manual recording. E4 also pointed out that ``\textit{the majority of semantic labels in narrations are aligned with corresponding animations.}"
And she suggested that semantic links between static visualization elements and static text narration segments can be established for narration-animation alignment and timeline management. 
E8 also mentioned this point and added that some authoring paradigms can be designed to help users effectively specify this semantic link.

The second insight pertains to animation design \reviseminor{(DC3)}. E3 suggested that visualization design knowledge can be leveraged to recommend suitable objects for animation as extensive research has been conducted on specific animation patterns for different visualization types~\cite{wang2022investigating}, such as bar growing, pie wheeling, line wiping, etc. 
Additionally, there are commonly used combinations of component-level animations. For example, when animating a pie chart, the sector wheels may be animated in conjunction with the corresponding legend flying in. Similarly, the bars of different groups may grow incrementally, and the titles and axes can fade in successively to enhance the overall presentation. These patterns can be effectively reused in animation design, especially making it easier to animate parallel content~\cite{Hullman2013a}.
Moreover, the relationship between animation effects and animated objects (\eg visual presence, visual attributes, and configuration summarized by Thompson et al.~\cite{thompson2020understanding}) can be utilized to recommend suitable animation effects. For example, given that most annotations are usually animated using visual presence effects, the tool can suggest a visual presence effect to the user when they select an annotation to apply animations.


\begin{figure*}[t]
  \centering
    \includegraphics[width=0.97\linewidth]{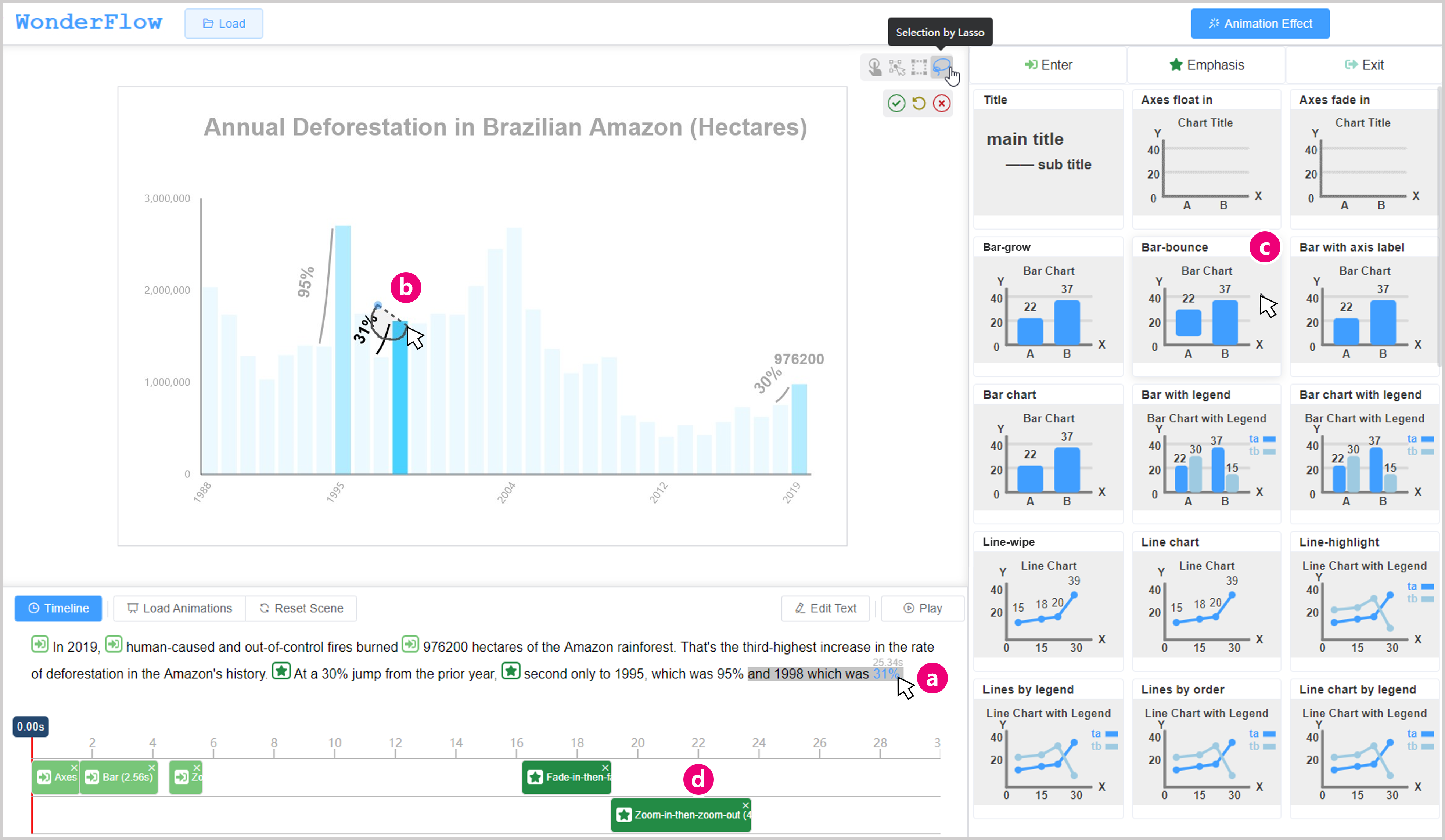}
    \caption{User interface of \toole. Users can first select the text phrases in the narration editor (a) and visual elements from the canvas (b) to form text-visual links. Then they can apply an animation preset selected in the animation effect panel (c) to the visual elements. \tool then generates a narration-animation pack on the timeline (d). 
    }
\label{fig:user interface}
  \vspace{-10px}
\end{figure*}

\subsection{Design Considerations}
From the formative study and previous work, we derive a set of design considerations to inform the design of our tool. 
The main goal of this work is to facilitate the design of narration-animation interplay and enhance data storytelling. In particular, we aim to enable non-designers like data analysts, journalists, and programmers to easily prototype data videos with narration-animation interplay.

\bpstart{DC1: Integrate narration and animation design for streamlined creation of data videos}
We want to integrate narration and animation editing features in a single tool, enabling users to easily create and arrange narration-animation interplay, and effortlessly edit their data videos. The separation of existing tools hinders the iterative design and preview of data storytelling~\cite{chen2018supporting, lee2015more}, particularly in the case of animation design and video editing tools. 
The system should not require users to customize narration content and animation designs in separate tools but offer an easy-to-use interface that enables them to seamlessly integrate the two.

\bpstart{DC2: Ease the creation of audio narrations} 
When creating audio narration for data videos, it can be tedious for authors to adjust the text content when designing animations and narrations together. With the advancement of text-to-speech technologies, there is an increasing percentage of generated voices online~\cite{cambre2020choice}. We aim to reduce the cost of creating audio narration by using text-to-speech techniques. The system can generate audio that can be directly used or served as a draft for the final data video. Alternatively, authors can record the audio, and the system can use an automatic speech recognition technique to ease text editing and time alignment when creating the videos.

\bpstart{DC3: Facilitate the animation design process} 
Creating an animation on visual components (\eg title, mark, and axis) involves designing, controlling, and setting animation parameters for multiple visual elements (\eg line, tick, arrow)~\cite{wang2022investigating, thompson2020understanding}. This process can be cumbersome. For example, changing the animation time and effect on one bar may require editing other bars in the chart as well. 
However, designers tend to use common design patterns when designing animations for common visualizations such as bar charts and pie charts\cite{amini2015understanding, kim2019designing}. We want to reuse these design patterns both to simplify this process and to enable non-designers to create animations with delicate details. To this end, the system should leverage the underlying visualization structure and retarget the animation design by encapsulating the animation design into reusable presets.


\bpstart{DC4: Support text-visualization linking and narration-animation synchronization} Narrations and animations are usually synchronized and they complement each other in data videos -- the narrations describe the informative context and data insights in a chart, while the annotations, marks or glyphs, and axes are usually animated during the narration~\cite{wang2022investigating}. 
When composing visual data stories, authors usually connect story pieces and sequence them into a script \cite{lee2015more}.
Visualizations and text are also semantically connected to enhance data communication~\cite{latif2021kori, sultanum2021leveraging}. If the system can leverage the semantic linking between text and visual components, the users no longer need to manually calculate the start time and duration of each animation along the timeline. 
We leverage semantic links between text and visual components and apply animations to them. In this way, the animations are designed with semantic meanings, telling stories vividly aligning with the text narration.   
}

%% file: sections/04-system.tex
\begin{figure*}[t]
  \centering
  \includegraphics[width=\linewidth]{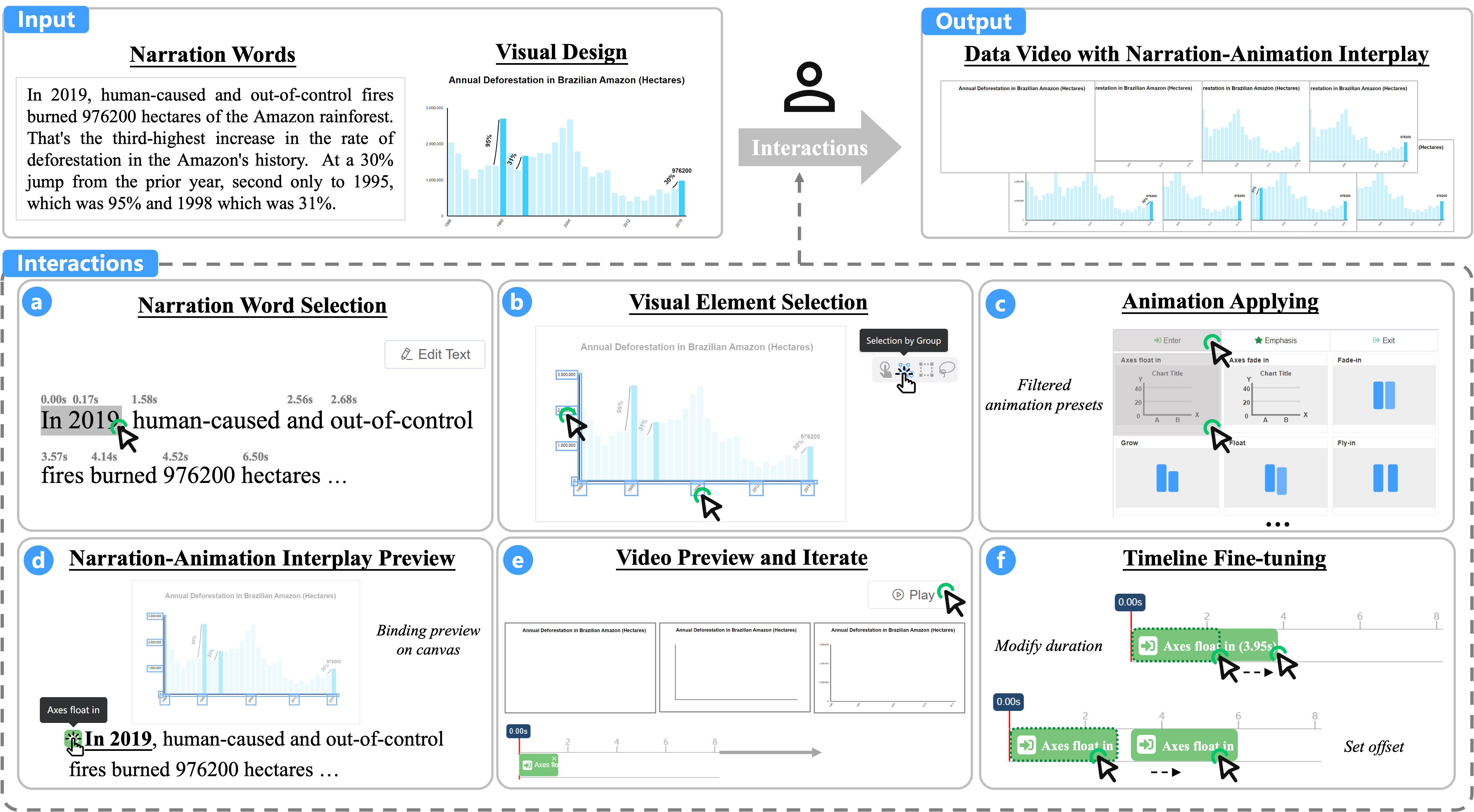}
  \caption{Interactions of \toole. The user can input the narration and \tool automatically generates the narration audio with timestamps (a). Users should first select narration words on the audio timeline (a) and then select visual elements on the canvas with various selection modes (b). Next, users can preview the filtered animation presets in the animation effect panel and apply an appropriate animation effect to the text-visual link (c). After that, hovering on the animation icon will show a preview of the created narration-animation binding (d). Then, users can click the ``Play'' button to compile the video for iterative preview (e). Finally, users can fine-tune the timeline (\eg duration and start) after preview (f) and iteratively design subsequent animations.}
  \label{fig:interface}  
  \vspace{-12px}
\end{figure*}

\begin{figure*}[t]
  \centering
  \includegraphics[width=\linewidth]{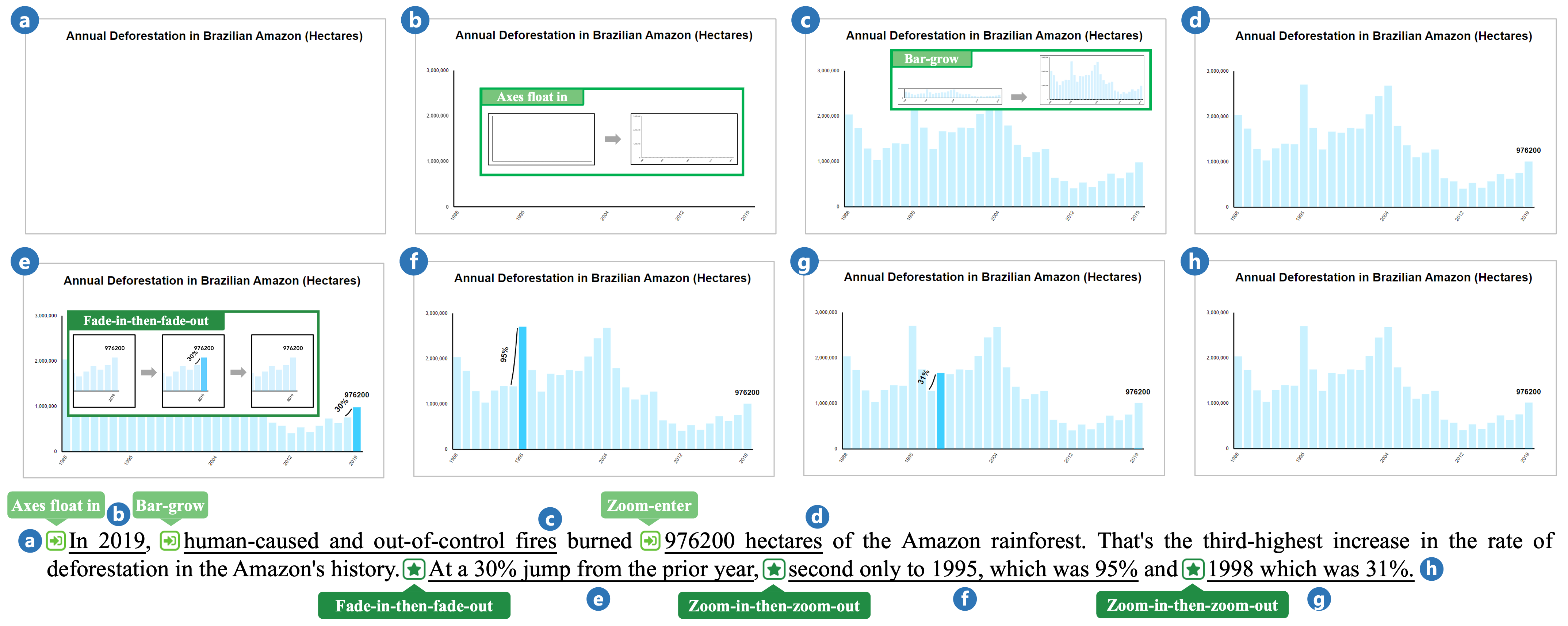}
  \caption{The generated data video in the use case. The bottom part shows the narration-animation linkings; the upper part is the video frames in specific timestamps, marked on the narration. The animation effects are as follows: at first, the canvas only presents a title, then the axes float in, followed by bars growing upwards from the bottom. Next, the annotation ``976200'' above the bar fades in. Finally, the three groups of annotations representing three major increases zoom in first and then zoom out in order (The data video can be found at \url{https://datavideos.github.io/WonderFlow/}).
  }
  \label{fig:use case}
    \vspace{-10px}
\end{figure*}

\section{\tool}
In this section, we first introduce the design pipeline behind \toole. Then we walk through a usage scenario to illustrate how users can interact with \toole. Next, we discuss structure-aware animations and introduce the narration-animation synthesis.
%

\subsection{Design Pipeline}\label{sec:pipeline}
We propose a pipeline of coupling narration to animation for easing the authoring process of a data video, as shown in Figure~\ref{fig:pipeline}. A data video incorporates multiple animation actions and an audio record.

Taking a static visual design and the corresponding text as input, users can conduct (a) text-visual linking through a graphical user interface \reviseminor{(DC4)}. To emphasize a specific text unit, the users first select the narration text and then link it to relevant visual elements. Then, the users (b) design narration-animation interplay by instantiating one or multiple animation actions with text-visual linking. In the back end, the system converted the text into \textit{audio narration} through text-to-speech techniques. The audio is stored in the form of a sequence of sentences, with the timestamps of each word \reviseminor{(DC2)}. The users can specify an animation action from the animation library \reviseminor{(DC3)}. An \textit{animation action} consists of predefined parameters, including the components that the animation effect is applied to, the time setting, and an animation effect. The selected narration and animation are temporally aligned \reviseminor{(DC4)}. Finally, the selected narrations and animations form multiple ``narration-animation'' pairs. The users can (c) preview the synthesized data videos and modify the text-visual linkings or animation actions, if needed \reviseminor{(DC1)}.
This pipeline bridges the gap between narration and animation by converting the mapping between temporal audio and temporal animation into a mapping between static text and static visual elements, providing an engaging experience to create data videos.


\input{sections/interface}

\begin{figure*}[t]
\centering
\includegraphics[width=\linewidth]{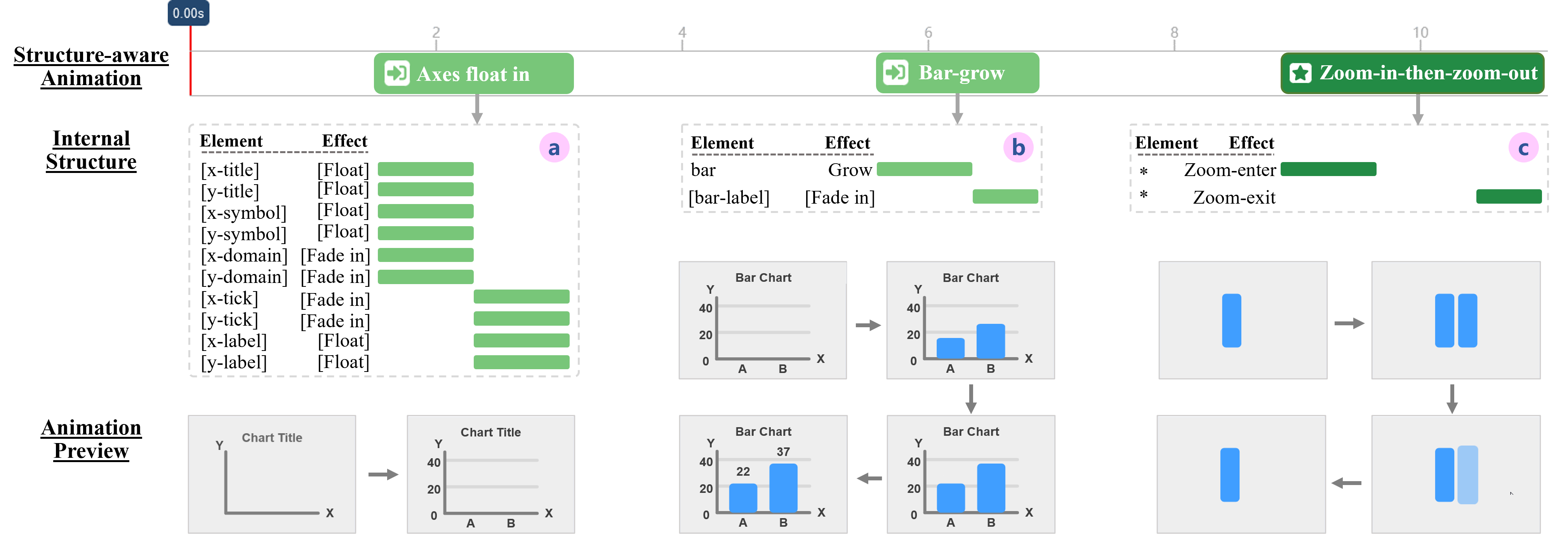}
\caption{Structure-aware animation presets. 
The upper part is three structure-aware animations created in the usage scenario (Figure~\ref{fig:use case}). The middle part is the internal structure of the presets. The square bracket ([]) in the animation card indicates soft constraints, which means that only if the elements exist, the corresponding animation effect will be applied. The asterisk (*) indicates that the animation can be applied to any visual element. The bottom part is the animation preview on the animation effect panel (Figure~\ref{fig:user interface}-c).
}
\label{fig:structure}
  \vspace{-10px}
\end{figure*}

\begin{figure}[t]
  \centering
  \includegraphics[width=0.95\linewidth]{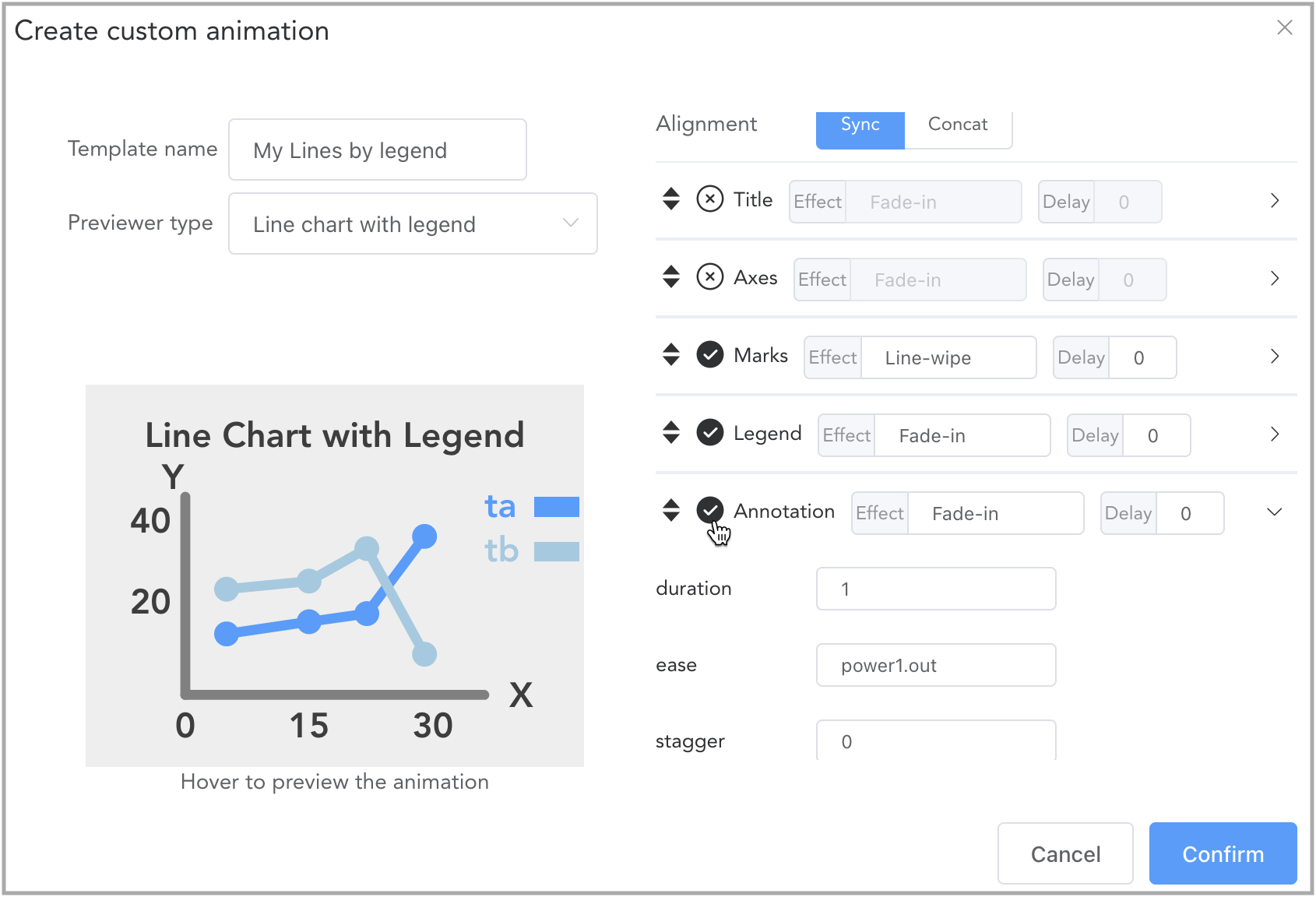}
  \caption{Animation Customization Panel.}
  \label{fig:Customization}
\vspace{-10px}
\end{figure}

\subsection{Structure-Aware Animation}
\tool links static visual elements to temporal animations. It achieves the flexibility of animation design by enabling users to upload their visual designs. In this subsection, we describe how the structure-aware animation effects in the animation library are applied \reviseminor{(DC3)}. We further enable animation customization, and simplify the interaction by animation filtering during users' visual element selection.

\subsubsection{Structure-Aware Animation Library}\label{sec:animation library}
A data video involves a set of animations on different visual elements that are related in the visualization structure (\eg a series of elements in axes, a set of marks, and a sector with a corresponding legend in the pie chart). We leverage this interrelationship to facilitate the animation creation process after linking the narration words and visual elements.

Specifically, we design a structure-aware animation library, which supports flexible customization of animations with a hierarchical granularity of visual components in a chart. For example, the x-axis is a chart component that consists of several graphical elements representing labels, ticks, and titles. When applying animations to the visualization components, our animation designs consider not only component type, but also the internal structure of the components. 
Figure~\ref{fig:structure} shows three example structure-aware animations used in the usage scenario and reveals the internal structure of the animation presets.
Each animation preset incorporates controls of multiple visual elements. For example, the ``Axes float in'' animation (Figure~\ref{fig:structure}-a) involves a set of effects for different axis-related elements. ``Bar-grow'' (Figure~\ref{fig:structure}-b) defines multiple bars growing synchronously, which requires that the selected visual elements should contain at least one bar element. If the selected elements include bar labels, they will fade in after the bars grow, which is a soft constraint. ``Zoom-in-then-zoom-out'' (Figure~\ref{fig:structure}-c) can be applied to any visual element. The applied elements will zoom in first and then zoom out after a certain interval.
Chart-level animations for the whole visualization are defined as a combination of lower-level animations for visual components. For example, a preset about the bar chart entering effect nests three sub-animations---``Title'', ``Axes float in'', and ``Bar-grow''---to show how the title, axes, and bar marks appear gradually. 

\revise{

\textbf{Structure-Aware SVG.}\label{SVG}
The structure-aware animations are applied to static visualizations with well-defined internal structures of visual components, in the format of \textit{Structure-Aware SVGs}.
Existing research systems use different file formats as input, such as data-enriched SVG for Canis \cite{ge2020canis} and CAST \cite{ge2021cast}, Vega-Lite for Gemini \cite{kim2021gemini}, or the file exported from Data Illustrator for Data Animator \cite{thompson2021data}. SVGs are widely used in visualization and animation authoring tools, but do not describe the structure information of the visualizations~\cite{wang2021animated}. Hence, we extend and revise the data-enriched SVGs \cite{ge2020canis} to include more common components in visualizations (\eg axes, legends, annotations) as structure-aware SVGs. Structure-aware SVGs have nested semantic labels of visualization components, following the structure of SVGs. The labels are stored as properties on the graphical elements. For example, for an x-axis, we label ``x-tick,'' ``x-title,'' ``x-label,'' etc. 
Inspired by existing SVG understanding technologies~\cite{poco2017reverse, Masson2023a}, we have implemented an automatic algorithm to convert SVGs to structure-aware SVGs inside \toole; but, we have found in practice that automatic methods may still produce errors when parsing visualizations with different design styles due to the complex SVG grammar.
To obtain high-quality structure-aware SVGs, we implemented a labeling tool to help users convert the SVG elements into structure-aware SVGs manually. 
We leave it as a future research direction to explore more efficient ways of combining manual efforts or improving the automatic algorithms to produce structure-aware SVGs. 



\textbf{Animation Presets.} The animation library enables good scalability and structure awareness. Currently, we have implemented 40 animation presets, covering common chart types (\eg pie chart, bar chart, line chart) and chart components (\eg coordinates, titles, marks). 
The animations are divided into three types: Enter, Exit, and Emphasis.
The entry animations include diverse structure-aware presets for different chart types, such as ``Pie-with-legend'' (sectors wheel in and the corresponding legends fly in), ``Lines-by-order'' (multiple lines wipe from left to right synchronously), ``Scatter-with-legend'' (the legends fade in first and then the scatters fade in staggered), etc.
Animations of the emphasis type are used to emphasize key points in the visualization. They usually call the user's attention by temporary appearance (\eg ``Fade-in-then-fade-out'',  ``Zoom-in-then-zoom-out'', and ``Show-arrow-label''), change in pattern (\eg ``Change-color''), dynamic effect (\eg ``Bar-bounce'' and ``Dynamic-number''), and view transformation (\eg ``Camera-zoom'' and ``Camera-move''). Among them, animations related to camera and move actions require users to input specific parameters, such as the ratio of ``Camera-zoom'', the specific coordinates of ``Move-to'', etc.
The exit effects include basic ``Zoom-exit'' and ``Fade-out''.

\textbf{Animation Customization.}
In addition to existing animation presets, \tool offers a customizable animation module that allows users to reuse patterns and personalize their own animation presets. 
Users can open the animation customization panel (Figure~\ref{fig:Customization}) to modify existing animations or create new ones. The customization panel includes the preview part showing the original animation template on the left, and the visual structure shown on the right.
For example, if users want to enhance the "Lines by legend" animation, they can configure parameters in the customization panel, such as adding annotations specifically for the visual structure of annotated lines by legend. This creates a new custom animation preset that expands the variety of animations with structure-aware effects and component types.
}

\subsubsection{Animation Filtering}\label{sec:animation filtering}
To help users rapidly find appropriate animations given the selected visual components, the system dynamically filters the animations applicable to the target components. 
As shown in Figure~\ref{fig:structure}, each animation contains multiple predefined sub-animations for a set of visual components. Define that animation is designed for the visual components $V := V_h \cup V_s$, where $V_h$ is the element set of hard constraints and $V_s$ is the element set of soft constraints (marked [ ]). $V_c$ is the set of selected elements on the visual canvas.
For each animation in the library, only if $V_s \subseteq V_c \subseteq V$, it will be presented on the animation panel for selection. 
In addition, users can further filter the animations on the interface based on the three animation behaviors, as shown in Figure~\ref{fig:user interface}-c and Figure~\ref{fig:interface}-c. When the number of preset animations grows, the system should consider built-in recommendation algorithms for users to find out the appropriate designs.

\subsection{Narration-Animation Synthesis}\label{sec:narration model}
Users can input the narration text into the narration editor, as shown in Figure~\ref{fig:user interface}-a. To generate the audio narration, \tool contains an audio synthesis module that leverages text-to-speech techniques to automatically transform the narration text into audio \reviseminor{(DC2)}. We use an online text-to-speech service~\cite{azurettsapi} with a performance comparable to natural speech ~\cite{Liu2021f}. It takes text as input, segments the text into sentences, and outputs the audio for each sentence and a list of timestamps for each word. The final generated audio narration provides a timeline for the data video. 
As such, the whole text input is decoupled into words with timestamps on the narration editor. Users can select a set of narration words to bind visual elements and further animations.

To coordinate animations with narration audio, the narration words and animations form narration-animation pairs and are arranged on a shared timeline \reviseminor{(DC4)}. This allows the animations to be mapped to word-level narrations, with the animations starting and ending at the same time as the selected text. Alternatively, users can use the default duration of the animation presets to avoid the animation effect being too long or too short. The default duration for the underlying animation is predefined in the animation library. For nested animations, the default value is the total length of the internal animation, and it speeds up or slows down if a specified duration is set. Additionally, users are allowed to fine-tune the time computation to arrange the audio and animation relationship, such as setting an offset of animations \reviseminor{(DC1)}.

%% file: sections/interface.tex

\subsection{Usage Scenario}
\subsubsection{User Interface}
\toole's user interface consists of (a) a narration editor, (b) a visual canvas, (c) an animation effect panel, and (d) a timeline (Figure~\ref{fig:user interface}).
Users can input the narration text of a data video in the narration editor and select a phrase with a mouse.
The visual canvas at the center displays a visualization. It has four selection modes that enable users to bind visual elements to the selected narration text \reviseminor{(DC4)}. In addition to commonly-used rectangle, single-click, and lasso selection, \tool supports a component selector based on the visualization structure, where users can select all the graphical elements in a component with one click. The animation effect panel on the right displays various animation presets of different granularity \reviseminor{(DC3)}. Hovering over the animation card shows a preview of the animation effect. After applying an animation to the selected narration text and visual elements, the narration-animation binding is temporally ordered on the timeline \reviseminor{(DC4)}. Users can further preview the video and perform fine-grained time adjustments, such as duration and trigger timestamp \reviseminor{(DC1)}. 

\subsubsection{Usage Scenario Walkthrough}

Imagine Nancy, a graduate student, is assigned a task to give a report on wildfires. After her investigation, she plans to use the example of bushfires that burned swaths of the Amazon rainforest to illustrate the great damage of wildfires. 
Nancy first collects the relevant data and creates a visualization. When she prepares her course presentations, she finds the static visualization is not so exciting or impressive enough to communicate the data insights she gained.
With little experience in animation and data video creation, she decided to use \tool to create a data video that can be embedded into her presentation.

Nancy opens \tool and uploads the prepared visualization, a bar chart with some annotations (Figure~\ref{fig:user interface}-b). Once she inputs the text script for the data video in the narration editor (Figure~\ref{fig:user interface}-a), \tool automatically generates the narration audio. She briefly checks the timestamps of the narration by moving the mouse over the words. 

With the narration text and visual design ready, Nancy starts to create animations.
To capture the audience's attention from the start, Nancy intends to apply an \textit{entry animation for the bar chart}.
Once she selects the starting narration words \sutt{In 2019} (Figure~\ref{fig:interface}-a), the visual canvas enters the selection state with a set of modes appearing on the right side (Figure~\ref{fig:interface}-b).
Nancy first chooses the selection mode by component, and clicks on an element on the x-axis. 
Then all elements in the x-axis group are selected, and enclosed by bounding boxes. 
She repeats similar operations to the y-axis group. 
Now that Nancy has built the semantic link between the narration words and visual elements, the corresponding animations for these selected elements are filtered out in the animation effect panel (Figure~\ref{fig:interface}-c). 
She briefly browses the animation previews and finds the ``Axes float in'' animation vivid, where the axis titles float in and axis domains fade in at the same time, followed by axis labels floating in and axis ticks fading in. Once Nancy selects the animation, the corresponding entry animation icon appears before the selected narration words on the timeline.
Hovering on the animation icon, \tool displays the narration-animation binding information (Figure~\ref{fig:interface}-d), including the selected text bolded and underlined in the narration editor, visual elements highlighted in the canvas, and the animation effect tooltip above the icon.
Now that the animation is specified, Nancy previews the data video by clicking the ``Play'' button (Figure~\ref{fig:interface}-e), and the corresponding video frames are shown in Figure~\ref{fig:use case} (a) and (b).

Nancy wants to \textit{show the bars} representing the annual deforestation in the Brazilian Amazon right after the appearance of coordinates with year labels on the x-axis and hectare ticks on the y-axis. So she binds the closely followed narration words \sutt{human-caused and out-of-control fires} with all the light-colored bar elements and applies a ``bar-grow'' animation. In the video preview, the bars grow upwards from the bottom (Figure~\ref{fig:use case}-c) at the start of \sutt{human-caused} and end with \sutt{fires} in the audio.

After the introduction of the bar chart, Nancy intends to \textit{emphasize the data insights} with animations on corresponding annotations. 
She expects the textual annotation ``976200'' to appear simultaneously with the narration \sutt{976200 hectares}.
By binding these two, she then applies a ``zoom-enter'' animation to make the text elements (``976200'' above the bar) enter (Figure~\ref{fig:use case}-d). 
Next, Nancy would like to emphasize the three major increases in deforestation rates in Amazon's history based on the final sentence in her narration script.
When browsing and previewing the animation library, she finds a ``zoom-in-then-zoom-out'' animation design interesting, with which the elements first zoom in and then zoom out after a certain duration. Before applying the animation, she links the words \sutt{At a 30\% jump from the prior year}, \sutt{second only to 1995, which was 95\%}, and \sutt{1998 which was 31\%} with the corresponding annotation elements (a text number, a dark-colored bar, and an arc), as selected in Figure~\ref{fig:user interface}-b. The corresponding data video frames are shown in Figure~\ref{fig:use case} (e), (f), (g), and (h). Ultimately, Nancy previews the whole video and fine-tunes the narration-animation packs (\eg duration and start) on the timeline (Figure~\ref{fig:interface}-f). She is happy with the expressive data video she quickly created with \toole.

%% file: sections/05-examples.tex
\begin{figure*}
  \centering
  \includegraphics[width=\linewidth]{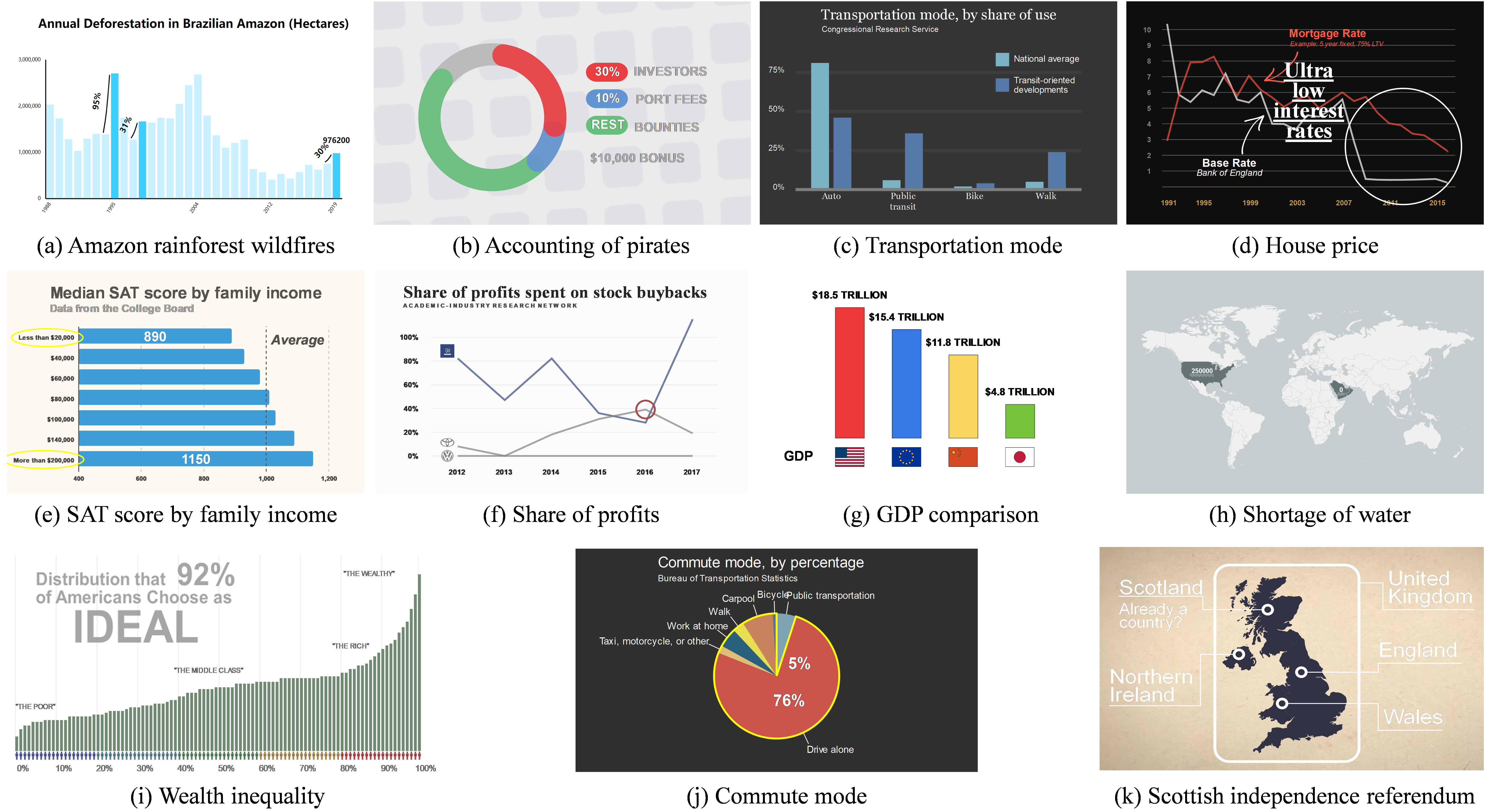}
  \caption{Visualizations in the example gallery derived from real-world data storytelling practices, covering diverse storytelling topics. 
  (More details about the gallery can be found at \url{https://datavideos.github.io/WonderFlow/}.)
  }
  \label{fig:example}
    \vspace{-10px}
\end{figure*}

\section{Evaluation}
We evaluate \tool via: 
(1) an example gallery to show the expressiveness of the generated data videos,
(2) a user study to access the system‘s usability,
(3) expert interviews to gather feedback from different perspectives,
and (4) an interaction-count comparison against PowerPoint.

\subsection{Example Gallery}
To demonstrate the expressiveness of \toole, we present a gallery of narration-enriched data videos\footnote{The whole example gallery can be found at \url{https://datavideos.github.io/WonderFlow/}.}, covering diverse visualizations and animation effects. 
The gallery involves various re-creations and variants of real-world data storytelling practices that sample the design space of narration-animation interplay, with the corresponding text scripts covering all semantic labels in narrations summarized in \cite{wang2022investigating}.  
Figure~\ref{fig:example} shows the visualizations used in our examples, which cover diverse chart types, annotations, and storytelling topics. Each example in the gallery includes a visual design, a piece of narration text, the original video link, the data video authoring process with \toole, and the final generated narration-enriched data video.




%% file: sections/06-user-study.tex
\subsection{User Study}\label{sec:user_study}
To evaluate the usefulness of \toole, we conducted an in-lab user study. 
\reviseminor{Inspired by existing evaluation strategies for authoring systems~\cite{Ren2019}}, the study includes both reproduction and open-ended design tasks with an aim to assess if non-experts could create data videos using \toole, and gain insights into how the tool improves the authoring experience for non-experts.  The reproduction study focuses more on evaluating the learnability and usability of the tool. The open-ended design tasks allow users to explore more features of the tool and try different design choices. This may provide more feedback about whether \tool can ease the creation process of data stories. 

\subsubsection{Participants}
We invited 10 participants (denoted as P1-P10, 4 females and 6 males, ranging from 23 to 40 years of age) to join our study. 
The participants are recruited by posting advertisements in a high-tech company. 
Various occupations were represented, including machine learning researchers, graduate students, UX designers, and data analysts. All participants had created data visualization for analyzing or presenting data in their daily work. However, their daily work does not involve animation creation, and they have no experience in creating data videos. On the other hand, all participants had created animations in Microsoft PowerPoint for presentation purposes. Only one participant used professional animation creation tools like Adobe After Effects.

\subsubsection{Procedure}
The study began with a brief introduction of the background and overall procedure of the study. We asked the participants to complete a pre-study background questionnaire and an interview. 
Then, we introduced the workflow, features, and interactions of \toole, and further demonstrated its usage by creating three example animations. 
The participants were encouraged to browse our animation library to understand the provided animation effects better. 
After that, they were asked to create animations by using the sample text and visualization to get familiar with the interface and operations. They can ask questions during the process. 
After training, the participants were required to complete a reproduction task and a free-form creation task. 

\bpstart{Reproduction task}
We asked the participants to replicate two reference videos (Amazon rainforest wildfires in Figure~\ref{fig:example}-a and accounting of pirates in Figure~\ref{fig:example}-b), where we provided the corresponding charts and textual narratives. 
Since some animation properties cannot always be clearly distinguishable by watching the reference videos, the animations are not required to be reproduced exactly the same. 
The resulting videos with a similar effect (``Fade-in'' vs. ``Float-in'') and approximate time properties (\ie start time and duration) are considered correct.
The two videos are 20 seconds long with five reference animations and 28 seconds long with six reference animations, respectively.


\bpstart{Free-form task}
After completing the two replication tasks, we asked the participants to create two videos freely, where we provided two static visualizations (transportation mode in Figure~\ref{fig:example}-c and house prices in Figure~\ref{fig:example}-d and the accompanying textual narratives. 
Before starting to create data videos, they were asked to familiarize the text and chart. They could also play the audio with static charts. They could ask questions about the data stories. They were also allowed to edit the text. During the creation, the participants were encouraged to think aloud, especially about the design considerations of animations. 

At the end of the study session, the participants filled out a questionnaire about \toole. The questionnaire is inspired by previous evaluation experience of authoring tools~\cite{Ren2019}. 
We further conducted a semi-structured interview with the participants to understand their design ideas about the data videos, and the experience of creating data videos with \toole. 
The interview lasted about 20 minutes. On average, the study lasted about 70 minutes.

\subsubsection{Results}
All participants were able to complete the reproduction and free-form tasks.
On average, the participants completed the two reproduction tasks in 6 and 7 minutes, respectively. The time to freely create the two data videos is 5 and 7 minutes, respectively, with 5 and 6 animations. 

\begin{figure}[t]
  \centering
  \includegraphics[width=\linewidth]{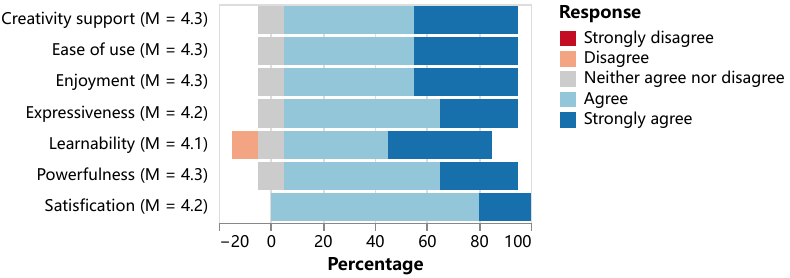}
  \caption{User ratings with a 5-point Likert scale.}
  \label{fig:questionaire}
\vspace{-10px}
\end{figure}

Overall the participants are positive about using \toole. Figure~\ref{fig:questionaire} shows the user ratings of the tool with a 5-point Likert scale.
The participants highly rated the satisfaction of \tool on a 1-5 Likert scale (\textit{M} = 4.20, \textit{SD} = .40, 5-Strongly agree, 1-Strongly disagree). They favored the usability of \tool in terms of ease of use (\textit{M} = 4.30, \textit{SD} = .64), learnability (\textit{M} = 4.10, \textit{SD} = .94), powerful functionalities (\textit{M} = 4.30, \textit{SD} = .60), enjoyment (\textit{M} = 4.30, \textit{SD} = .64), expressive animation design (\textit{M} = 4.20, \textit{SD} = .60), and creativity support (\textit{M} = 4.30, \textit{SD} = .64).

All the participants found \tool useful for creating data videos. In particular, they found \tool helpful to animate charts quickly in a single platform \reviseminor{(DC1)}. P5 said, ``\textit{it is simple to add or change an animation.}'' 
The participants found it enjoyable to create short data videos and watch the videos they had created, even when they had little experience in creating motion graphics. P3 told us, ``\textit{It is my first time creating this kind of video. I imagined it should be made by professional animators. It is surprising that the videos can be made by myself. I really like the outcomes.}'' 
The participants said they would use the tool in the future when they ``\textit{want to present data results to important people and leave a deep impression on them}'' (P6). The participants also felt it increasingly common to present results in the form of video because ``\textit{using a video form seems an effective way of sharing insights and can be done remotely and asynchronously}'' (P2).

The participants found \tool a good way to integrate narrations and animations \reviseminor{(DC4)}. They could easily understand the temporal relations between animations and narrations. One participant (P4) mentioned, ``\textit{I no longer need to set the time (manually) by myself. It is very convenient.}'' 
Another participant (P8) also commented that ``\textit{The text-visual linking interaction is easy to learn and use, which is an interesting and useful design.}''
From our observation, the ease of creating these narration-animation interplay has encouraged the participants to create fine-grained animation designs, \eg showing the axes and bars when mentioned in the audio, instead of showing all the visuals at once. 
This also further encouraged more design exploration. The participants enjoyed the design process of trial-and-error, while other tools may introduce too high barriers or more time and effort to search for better results.
In addition, the participants (P4, P6, P7, P9) also commended the integrated audio-generation techniques as they often made slips of the tongue when recording and had to re-record, which was tedious \reviseminor{(DC2)}. P6 mentioned, ``\textit{when using the system (\toole), there is no need to care about the audio as it is automatically generated, and the effect is satisfactory.}''
We observed the participants sometimes made mistakes when conducting text-visual linking. They may start from selecting visual elements and applying animations before linking with text selection, and then correct their operations. We suspect this is a habit caused by familiarity with using presentation tools such as MS PowerPoint or Keynotes. 

The participants felt the structure-aware presets were useful to facilitate the animation binding process \reviseminor{(DC3)}.
P9 mentioned, ``\textit{previously, I needed to add animations to multiple visual elements in tools like PowerPoint. Now, I only need to use one animation effect, which contains a sequence of sub-animations applied to multiple elements.}” In comparing \tool with animation design tools, P4 commented, ``\textit{I no longer need to set animation effects frame by frame. It is easier than using AE [Adobe After Effects] to craft animations.}” The participants also found the animation library useful to help them explore new animation designs. P3 said, ``\textit{I enjoy trying different kinds of animation effects to see what I can achieve.}” 
While useful, we found that the participants still needed time to familiarize all the animations. It took time for them to preview and understand the animation effects. 
The advantage of the animation library may increase when the users get more familiar with the expected effects on their charts. Different from other participants, P10 requested to design her own structure-aware animations, ``\textit{I would like to refine the animations, set the parameters by myself, and save them for my future usage.}''

When asked about future improvements to the tool, P1 required more control over the dependency among animations (\eg \textit{begin after}, \textit{end together}, \textit{begin together}, or \textit{start at an exact time point}) to further ease the animation design. 
P6 mentioned he wanted to have more suggestions on the animation types. For example, the system may suggest a grow-out animation for a column chart; the system could also help the users avoid using multiple enter effects for one object, or give a warning when the users insist on doing so. 
The participants (P7, P10) further mentioned that it would be better to have in-progress previews at any time point, especially when there are many elements and animation effects to manage on the canvas.


\revise{
\subsection{Expert Interview}
In this interview, four expert users were invited to provide feedback on \tool for creating data videos. Two motion graphic designers (D1 and D2) have 4 and 6 years of experience, respectively, primarily using Adobe After Effects for motion graphic video creation, and two visualization researchers (V1 and V2) have 5 and 6 years of experience researching visualization tools and techniques, respectively. Both researchers have published visualization papers at top conferences (\eg IEEE VIS and ACM CHI).

The four experts were shown six data videos randomly selected from the novice-created data video results, including two from the reproduction study and four from the free design study. Both designers (D1 and D2) found the videos satisfactory and commented that no further improvements were needed. D2 mentioned that if the videos were incorporated into longer ones, he would adapt the design and animation styles. The four experts were then introduced to the functions of \tool and asked to try the tool out for themselves. The designers (D1 and D2) found the tool to be user-friendly and opined that the use of Adobe After Effects to create comparable data videos is a complex process. 
They also recommended integrating the functionalities of \tool, such as element selection and animation recommendations, into Adobe After Effects.

The two researchers specializing in visualization (V1 and V2) had different perspectives on the usefulness of \toole. 
V1 noted that the tool enables the quick and easy creation of a data video, starting from writing the narration all the way to explaining the insights from the chart. V2 proposed that the tool could be more valuable if integrated into business intelligence tools like Tableau and Power BI. V2 explained, ``\textit{The automated insight mining techniques could be combined to automatically discover interesting stories and generate narrations. I can imagine I would only need a few clicks to generate a video.}''

The experts further provided valuable feedback on the potential
application scenarios. 
They were all positive about the topic and recognize the importance of the narration-animation interplay.
V2 commented, ``\textit{compared with fine-grained controls in professional video creation tools like AE (Adobe After Effects), the design pipeline behind \tool provides an in-the-middle layer of abstraction.}''
They all commented that \tool with easy-to-learn interactions can effectively facilitate the creation of narration-enriched data videos, especially for novices. Moreover, D1 said that \tool can also be used to prototype data videos for them (professional designers). 
In addition, V1 said that the design pipeline can also be extended to support more design types such as motion graphics and infographics.
Besides, when asked about the limitations, 
three participants (D1, D2, and E1) indicated that the usability of the tool could be improved, which is consistent with the feedback from the in-lab study, which will be further discussed in Section~\ref{sec:discussion}.

\begin{table}[t]
\centering

\caption{Number of interactions for authoring animations in Figure~\ref{fig:example} with \tool and PowerPoint, respectively.}
\setlength{\tabcolsep}{1mm}{
\begin{tabular}{c|c|c|ccc}
\toprule
 && \multicolumn{4}{c}{WonderFlow/PowerPoint} \\
\midrule
Example& \#Element & \#Animation& \#Click & \#Brush & \#Keyboard  \\
\midrule
(a) & 55& 6/10 (-4)& 17/53(-36) & 3/3 (0)  & 0/13 (-13) \\
(b) & 14& 5/8 (-3)& 15/21 (-6)& 3/5 (-2) & 0/0 (0) \\
(c) & 27& 5/10 (-5)& 14/24 (-10)& 1/4 (-3) & 0/1 (-1)\\
(d) & 40& 7/10 (-3)& 26/39 (-13)& 0/4 (-4) & 0/5 (-5)\\
(e) & 33& 7/14 (-7)& 24/42 (-18)& 0/4 (-4) & 0/4 (-4)\\
(f) & 31& 6/12 (-6)& 24/37 (-13)& 1/4 (-3) & 0/3 (-3)\\
(g) & 13& 5/9 (-4)& 11/19 (-8)& 5/5 (0)  & 2/4 (-2) \\
(h) & 221& 7/7 (0) & 20/26 (-6)& 3/3 (0)  & 4/2 (2)\\
(i) & 230& 5/10 (-5)& 11/29 (-18)& 5/8 (-3) & 0/5 (-5)\\
(j) & 27& 7/9 (-2)& 21/35 (-14)& 1/4 (-3) & 0/6 (-6) \\
(k) & 21& 10/10 (0)& 28/46 (-18)& 5/5 (0)  & 0/8 (-8)  \\
\bottomrule
\end{tabular}}
\label{tab: comapre}
 \vspace{-10px}
\end{table}

\subsection{Comparison}
We also compared \tool with the existing data video creation workflow with Microsoft PowerPoint. After loading visual elements, PowerPoint allows users to animate elements, record audio narration, align animations and narrations, and export to a video. An ideal approach is to conduct a comparative user study, but the two tools differ greatly in design architecture, key features, interface operations, underlying implementation logic, etc. 
Therefore, inspired by the keystroke-level model~\cite{Card1980}, we chose to compare the number of user interactions as an alternative assessment of their respective complexity. 

For the animation part, we animated the 11 charts in Figure~\ref{fig:example} with \tool and PowerPoint, respectively. The animation is considered successful if the animation of visual elements can be implemented with a reasonable design and focus on the correctness of semantic communication. In addition, we disregarded the fine-tuning process and focused only on the minimum set of interactions. Table~\ref{tab: comapre} shows the minimum number of necessary animations, and the number of required Click, Brush, and Keyboard interactions. They can be considered as a general impression of the interaction complexity. From the table, we found that \tool requires relatively fewer low-level interactions.

In other aspects, \tool and PowerPoint differ significantly in the design logic, so here we only discuss the comparison qualitatively. Regarding narration recording, \tool directly calls the TTS service to generate and update audio at any time, while PowerPoint requires manual recording or the use of third-party tools to generate voice files and then import them into the system, which is cumbersome. Regarding audio and animation synchronization, the text-visual linking paradigm designed in \tool allows users to first select a narrative segment and then bind it to semantic-related visual elements and animations. The system will automatically achieve alignment on the timeline, which will introduce only one additional Brush interaction. However, PowerPoint requires users to listen to the audio repeatedly in the "recording" function and trigger the corresponding animation at the right time, which is very inconvenient. 
Additionally, in \tool, users can easily adjust the timeline without the need to re-listen to audio or realign animations like in PowerPoint. Moreover, \tool provides real-time preview and fast-forward capabilities for specific animations, whereas PowerPoint only allows preview at the slide page level.

Based on this data and analysis, we concluded that \tool has a clear advantage over PowerPoint in terms of interaction complexity when creating data videos.

}

%% file: sections/07-discussions.tex
\section{Discussion}\label{sec:discussion}
\revise{
This section discusses lessons we learned during the design of \toole, future work, and limitations.

\subsection{Lessons Learned}
\bpstart{Enabling multiple interaction pipelines for creating narration-animation interplay}
\tool is designed with a new pipeline, asking the users to link text and visuals first, and then bind the animations to text-visual links \reviseminor{(DC4)}.
Users can easily select the animations endowed with semantic meanings without explicitly documenting the intents behind animation designs \reviseminor{(DC3)}. However, we also find that users from different design backgrounds require multiple interaction pipelines, besides text-visual linking, to more flexibly meet their design requirements.
The participants (P1, P3, P5, P7) mentioned the interactions with the text guided them to design the animations from the perspective of narration-animation interplay. 
If they can select visual elements first and then apply animation effects to the elements, similar to design tools like MS PowerPoint, it could make the process more intuitive and reduce the learning curve. Another possible way is to build an example-driven system and encourage users to start by selecting animation actions from the library or providing a reference~\cite{Shen2022b, Xie2023a}. Correspondingly, the structure-aware animations should be directly applied to the related components.
This might be useful to reduce the efforts of exploring animation effects from the library, but may introduce ambiguity when multiple components are applicable on the canvas. Future research may start by comparing other potential pipelines or provide more flexibility by enabling users to switch among pipelines freely.

\bpstart{Allowing diverse temporal relations in narration-animation interplay}
Leveraging the commonness of echoes in narration-animation interplay in data videos~\cite{wang2022investigating}, the pipeline of \tool infers the temporal interplay of narrations and animations from text-visual links. From the study, the participants felt this method could save lots of effort for time-aligning narrations and animations that are semantically related. They no longer need to manually synchronize timing for audio narration and animation \reviseminor{(DC4)}. However, there are cases where animations are not closely related to the semantics in narrations. For example, the entering of the coordinates may start with the video, but is not closely related to semantics in the narrations; some animations may temporally and logically depend on other animations (\eg text annotations shown right after the arrows); some animations may be designed purely for embellishments and do not have semantic meanings. While they are not the majority according to the previous study~\cite{wang2022investigating}, future work should consider supporting these cases for an improved experience of animation design in data videos. 

\bpstart{Easing animation design with visualization structure-aware animation presets}
Different from the previous composition of animation libraries in general tools like MS PowerPoint and Adobe After Effects where users need to apply animations element by element,
\tool supports structure-aware animation presets.
The structure-aware animation presets are tailored for data-related videos \reviseminor{(DC3)}. The insight is that animations designed for a certain type of chart may be applied across chart designs by leveraging the internal structure of common visualization (\eg axes wipe out and bars grow up in bar charts). 
Since all the users found the animation library enough for them to create animations, we envision more animations to be supported for more chart types and components. Moreover, the animation library can also be extended to support common visual groups, such as bullet items. Targeting novice users, low-level parameters for the animations are not revealed to users. One participant (P9) mentioned that she enjoyed the process of exploring the animation effect to make it personal. It will be interesting to explore how we can facilitate novice users to inject their animation preferences to create their own animation presets. For example, users can use natural language to describe animations for generation and updating~\cite{vistalk}, \eg ``wipe in the axes and then grow out the lines one by one''. Another direction is to design multi-modal interactions to specify in-detail temporal dependencies within an animation. 

\subsection{Future Work}
\bpstart{Extending to leverage other tools' authoring capabilities, also in other domains}
\tool is currently implemented as a standalone application, separating chart creation from data video creation. Converting SVGs to structure-aware SVGs, \tool can leverage the design capabilities of other tools to create chart designs. We envision it can also be useful as an add-in for other tools in different user scenarios. For example, \tool can be implemented as extensions for interactive data article tools \cite{latif2021kori, sultanum2021leveraging} to convert data articles to data videos. With these tools, users can establish semantic links between text and visuals automatically or semi-automatically, which is compatible with our design pipeline and saves manual efforts.
Additionally, \tool can be implemented as extensions for visualization authoring tools or data analysis tools (\eg computational notebooks) to bridge the gap between data analysis and communication~\cite{Li2023,inkinsight}. The internal structures of visualizations are naturally maintained. The analysis results can be more easily explained and shared with narrations and animations. Future research can be conducted on whether data analysts are interested in sharing data stories with \tool in their daily work. 
\tool can also be extended to support more data video features, such as narrative structure~\cite{Yang2022a}, cinematic effects~\cite{Xu2023b}, and emotion expression~\cite{Xie2023}.
\reviseminor{Our technology can further be extended to benefit other domains as well, such as generating educational content for courses and producing travel promotional materials.}

\bpstart{Combining more automation techniques to ease the process} 
Our primary focus in this paper is on the novel process of authoring the narration-animation interplay. 
More automated techniques can be applied to reduce manual data video creation efforts. 
\reviseminor{For example, the text-visual linking and animation selection steps can be automated with the help of large language models (LLMs) and constraint programming~\cite{data_player}.
After collecting users' interaction records of text-visual linking, machine learning models can be trained to learn semantic linking patterns between narration and visual elements. Users only need to further specify narration-animation interplay. }
In addition, as discussed in Section \ref{SVG}, future advances in the reliability of reverse-engineering techniques can ease the conversion from SVGs to structure-aware SVGs~\cite{Ying2023, Dou2024}. Furthermore, it will be worth exploring techniques of parsing animations from data videos and extracting reusable animations from existing ones. 
In our pipeline, we use the text-to-speech technique to generate audio. In the future, we can leverage speech recognition to enable users to input their pre-recorded audio to make the final effect more personalized. Additionally, human speech contains acoustic signals (\eg pitch, pause, tone, etc.) to convey sentiments, which might be useful for the system to recommend animations of different styles (\eg lively, calm, professional, etc.). 
\reviseminor{\tool requires users to write narrations and design visualizations by themselves. Alternatively, starting from data, LLMs can be used to mine data insights, generate visualizations, and write text narrations based on the analytic task.}
In addition, it would be interesting to investigate at which stages of data video generation users would like to be assisted by AI and how they expect to collaborate with AI~\cite{Li2023a, Shi2023, Chen2022, Li2023c}. }

\subsection{Limitations}
Our study participants are limited to novice users for animations. Though their daily job does not involve animation design, they all have experience in creating animations in Microsoft PowerPoint or Apple Keynote. Our work can be complemented with future studies on participants with diverse backgrounds in creating animations (\eg people without any animation creation experiences). We note that creators with a certain type of disabilities (\eg hearing impairment) may benefit from our approach that converts text-visual linking to narration-animation binding. However, people with other disabilities (\eg vision or motor impairment) may have difficulties in using \tool to create data videos. Further research is needed to make the authoring experience more accessible to them~\cite{marriott2021inclusive}. 

In addition, the time for getting familiar with the tool during the reproduction and free-form tasks is limited. This research lacks evaluation on how long-term real-world usage may affect users' impressions of the tool. In the in-lab study, we prepared in advance the input text and visualization for the users to save time. The users are presumed to be able to create charts and text descriptions and understand the components of the charts. Future research could assess if people are able to create expressive data videos with \tool from scratch, and how people may integrate data video creation into their data analytics workflow. Besides, using words as time anchors for animation design has reduced the time cost, but the animation designs largely depend on the text input. The time granularity is limited to the start time of each word. Although users can set offset for the animations, this may take more effort and discourage users from fine-tuning time settings. This may be addressed by complementing more features of creating animations in our timeline view in the system. 
\reviseminor{Furthermore, the current design of the structure-aware animation library is primarily focused on common chart types, as described in Section \ref{sec:animation library}. However, there is potential for further expansion to support a broader range of visualization types, such as semi-structured data and text data representations~\cite{Shen2022, Xie2023a, Xie2023}. Additionally, the library can be extended to accommodate multi-chart scenarios, enabling transition animations between different visualizations~\cite{kim2021gemini,kim2021gemini2}.}

%% file: sections/08-conclusion.tex
\section{Conclusion}
We presented an authoring tool for narration-centric design of animated data videos. We proposed a design pipeline to enable the interplay of narrations and animations. We used static text-visual correspondence to ease the temporal alignment of narration and animation.
Based on the pipeline, we further designed and implemented an authoring tool \tool with a novel user interface and interaction design -- users first specify text-visual links, then apply an animation effect chosen from the structure-aware animation library, and finally preview to iterate the designs. To assess our approach, we presented an example gallery of real-world storytelling practice and conducted a user study and expert interviews to evaluate the learnability and usability. 
We hope that our study can inspire more interesting research on narration-animation interplay in the future, such as exploring different authoring pipelines and developing more intelligent and automatic approaches to support more complex data stories.

\section{Acknowledgement}
The authors would like to thank the anonymous reviewers for their valuable feedback, Mingxuan Fan, Xingyun Chang, and Yuheng Wu for discussions and initial prototyping related to this work during their internship.
This work was funded in part by the Institute of Information and Communications Technology Planning and Evaluation (IITP) Grant funded by the Korean Government (MSIT), Artificial Intelligence Graduate School Program, Yonsei University, under Grant RS-2020-II201361.